\documentclass[aps, prd, amsmath, floats, floatfix, twocolumn, nofootinbib, superscriptaddress, showpacs]{revtex4-1}
\usepackage{graphicx}
\usepackage{color}
\usepackage{latexsym}
\usepackage{amsfonts}

\newcommand{\be}{\begin{eqnarray}}
\newcommand{\ee}{\end{eqnarray}}

\begin{document}

\title{Testing Einstein-dilaton-Gauss-Bonnet gravity with the reflection spectrum of accreting black holes}

\author{Hao Zhang}
\affiliation{Center for Field Theory and Particle Physics and Department of Physics, Fudan University, 200433 Shanghai, China}

\author{Menglei Zhou}
\affiliation{Center for Field Theory and Particle Physics and Department of Physics, Fudan University, 200433 Shanghai, China}

\author{Cosimo Bambi}
\email[Corresponding author: ]{bambi@fudan.edu.cn}
\affiliation{Center for Field Theory and Particle Physics and Department of Physics, Fudan University, 200433 Shanghai, China}
\affiliation{Theoretical Astrophysics, Eberhard-Karls Universit\"at T\"ubingen, 72076 T\"ubingen, Germany}

\author{Burkhard Kleihaus}
\affiliation{Institut f\"ur Physik, Carl von Ossietzky Universit\"at Oldenburg, 26111 Oldenburg, Germany}

\author{Jutta Kunz}
\affiliation{Institut f\"ur Physik, Carl von Ossietzky Universit\"at Oldenburg, 26111 Oldenburg, Germany}

\author{Eugen Radu}
\affiliation{Departamento de F\'isica da Universidade de Aveiro and CIDMA,
 Campus de Santiago, 3810-183 Aveiro, Portugal}

\date{\today}

\begin{abstract}
Einstein-dilaton-Gauss-Bonnet gravity is a theoretically well-motivated alternative theory of gravity emerging as a low-energy 4-dimensional model from heterotic string theory. Its rotating black hole solutions are known numerically and can have macroscopic deviations from the Kerr black holes of Einstein's gravity. Einstein-dilaton-Gauss-Bonnet gravity can thus be tested with observations of astrophysical black holes. In the present paper, we simulate observations of the reflection spectrum of thin accretion disks with present and future X-ray facilities to understand whether X-ray reflection spectroscopy can distinguish the black holes in Einstein-dilaton-Gauss-Bonnet gravity from those in Einstein's gravity. We find that this is definitively out of reach for present X-ray missions, but it may be achieved with the next generation of facilities.
\end{abstract}

\maketitle


\section{Introduction}

Einstein's gravity is our current framework for the description of the gravitational field and the chrono-geometrical structure of the spacetime and has passed a large number of observational tests. However, it has been mainly tested in weak gravitational fields, with experiments in the Solar System and radio observations of binary pulsars~\cite{will}. A number of alternative theories of gravity have the same behavior as Einstein's gravity in the weak field regime and present observable deviations only when gravity becomes strong~\cite{berti}. In this context, astrophysical black holes are the best laboratory to test strong gravity.

There are two main lines of research to test the nature of astrophysical black holes: the study of the properties of the electromagnetic radiation emitted by gas or stars orbiting these objects~\cite{r1,r2} or the analysis of the gravitational wave signal emitted by a system with a black hole~\cite{r3,r4}. Tests with electromagnetic radiation include, but are not limited to, the study of the thermal spectrum of thin accretion disks~\cite{cfm1,cfm2,cfm3,cfm4}, the analysis of the reflection spectrum of thin disks~\cite{i1,i2,i3,i4}, the measurements of the frequencies of quasi-periodic oscillations~\cite{qpo1,qpo2,qpo3,qpo4}, and the possible future detection of black hole shadows~\cite{s1,s2,s3,s3.5,s4,s5,s6}. Among these techniques, X-ray reflection spectroscopy is the only one that can be already used to test astrophysical black holes and promise to be able to provide stringent constraints with the next generation of X-ray facilities~\cite{next1,next2,next3}.

The approaches to test astrophysical black holes include on the one hand
model-independent tests. These employ a parametrized metric in which possible deviations from the Kerr solution are described by a number of deformation parameters, see, for instance, Refs.~\cite{p1,p2,p3,p4}. This strategy is a reminiscent of the PPN formalism to test the Schwarzschild solution in the weak field limit with Solar System experiments. However, in the case of tests in the strong gravity regime it is not possible to write the most general expression for the metric with a well-defined hierarchical structure.

On the other hand, an alternative approach is to test a specific theory and check whether observational data prefer the Kerr black holes of Einstein's gravity or the non-Kerr black holes of the alternative theory of gravity under consideration. Unfortunately, this approach can be rarely adopted because rotating black hole solutions are very difficult to obtain. In alternative theories of gravity, we often know the non-rotating solutions, sometimes we know the rotating solutions in the slow-rotation approximation, but only in quite exceptional cases we know the complete solutions valid even for fast-rotating black holes. This is a problem, because astrophysical objects have naturally a non-vanishing angular momentum and fast-rotating black holes are the most suitable sources for testing strong gravity, as the inner edge of the disk gets closer to the compact object, maximizing the relativistic effects in the electromagnetic spectrum of the source.

The aim of this paper is to present a preliminary study on the possibility of distinguishing the Kerr black holes in Einstein's gravity from the black holes in Einstein-dilaton-Gauss-Bonnet (EdGB) gravity with present and future X-ray missions from the analysis of the disk's reflection spectrum. 
Black holes in EdGB gravity are quite a special case, however,
since besides the static \cite{bh1}
and slowly rotating black holes \cite{bh2,bh3}
also the rapidly rotating solutions are known numerically~\cite{bh4,bh5}. 
We can thus expect to test this theory and constrain its fundamental parameters from astrophysical observations of black holes. 
Previous attempts along this line
include the analysis of quasi-normal modes \cite{res1}
and the investigation of the shadow of EdGB black holes \cite{roman,res2}.

Here we consider the reflection spectrum of accreting black holes. We do not analyze real data, but we simply study the constraining power of possible observations with simulations. We simulate observations with NuSTAR (current X-ray mission) and eXTP~\cite{extp} (next generation of X-ray facilities) of a bright black hole binary. We find that NuSTAR cannot distinguish a Kerr black hole from a black hole in EdGB gravity. On the contrary, eXTP seems to be able to do it if we have the correct astrophysical model.

The content of the present paper is as follows. In Sections~\ref{s-edgb}, we briefly review the rotating black hole solutions in EdGB gravity and we choose a set of numerical metrics to be studied in the sections after. In Sections~\ref{s-iron}, we describe our astrophysical model and the main properties of the reflection spectrum of thin accretion disks. In Section~\ref{s-sim}, we present our simulations with NuSTAR and LAD/eXTP. Section~\ref{s-dis} is devoted to the discussion of our results. Summary and conclusions are in Section~\ref{s-con}. Throughout the paper we employ natural units in which $c = G_{\rm N} = \hbar = 1$ and a metric with signature $(-+++)$.

\section{Black holes in E\lowercase{d}GB gravity \label{s-edgb}}

EdGB gravity is one of the simplest string-inspired 4-dimensional models with higher curvature terms
and also can be seen as a
particular case of Horndeski gravity \cite{berti}. 
The field equations are still of second order and the theory is ghost-free. The action reads 
\be
\hspace{-0.5cm}
S = \frac{1}{16\pi} \int d^4x \sqrt{-g} \left[ R - \frac{1}{2} \left(\partial_\mu \phi \right)^2
+ \alpha e^{-\gamma\phi} R_{\rm GB}^2 \right] \, ,
\ee
where $\phi$ is the dilaton, $\alpha$ and $\gamma$ are coupling constants, and $R_{\rm GB}^2$ 
is the Gauss-Bonnet term
\be
R_{\rm GB}^2 = R^{\mu\nu\rho\sigma} R_{\mu\nu\rho\sigma}
- 4 R^{\mu\nu} R_{\mu\nu} + R^2 \, .
\ee

Rotating black hole solutions can be obtained employing  a metric ansatz
in quasi-isotropic coordinates
\be
ds^2 &=& - f dt^2 + \frac{m}{f} \left( dr^2 + r^2 d\theta^2 \right) \nonumber\\
&& + \frac{l}{f} r^2 \sin^2\theta \left( d\phi - \frac{\omega}{r} dt\right)^2 \, ,
\ee
where the metric functions $f$, $m$, $l$, and $\omega$ 
as well as the dilaton function $\phi$
depend on the coordinates $r$ and $\theta$ only. The boundary conditions at the event horizon $r = r_{\rm H}$ are
\be
&& f \big|_{r = r_{\rm H}} = m \big|_{r = r_{\rm H}} = l \big|_{r = r_{\rm H}} = 0 \, , \nonumber\\
&& \omega \big|_{r = r_{\rm H}} = \Omega_{\rm H} r_{\rm H} \, , \quad
\partial_r \phi \big|_{r = r_{\rm H}} = 0 \, ,
\ee 
where $\Omega_{\rm H}$ is the angular velocity of the horizon. The boundary conditions at infinity are
\be
&& f \big|_{r = \infty} = m \big|_{r = \infty} = l \big|_{r = \infty} = 1 \, , \nonumber\\
&& \omega \big|_{r = \infty} = 0 \, , \quad
\phi \big|_{r = \infty} = 0 \, ,
\ee
in order to have an asymptotically-flat spacetime with a vanishing dilaton at infinity.
Furthermore, axial symmetry, reflection symmetry, and regularity require the following boundary conditions
\be
&& \partial_\theta f \big|_{\theta = 0,\pi/2} = \partial_\theta l \big|_{\theta = 0,\pi/2} =
\partial_\theta m \big|_{\theta = 0,\pi/2} = 0 \, , \nonumber\\
&& \partial_\theta \omega \big|_{\theta = 0,\pi/2} = 0 \, , \quad
\partial_\theta \phi \big|_{\theta = 0,\pi/2} = 0 \, .
\ee
The absence of conical singularities implies
\be
m \big|_{\theta = 0} = l \big|_{\theta = 0} \, .
\ee

Rotating black hole solutions are obtained numerically solving a set of five second-order, coupled, non-linear, partial differential equations for the functions $f$, $m$, $l$, $\omega$, and $\phi$ imposing the above boundary conditions~\cite{bh4,bh5}. Besides the coupling constants $\alpha$ and $\gamma$,
the input parameters are the horizon angular velocity $\Omega_{\rm H}$ and the
horizon radius $r_{\rm H}$. At the end of the integration the horizon
area $A_{\rm H}$ is obtained from the metric at the horizon,
while the mass $M$, the spin angular momentum $J$, 
and the dilaton charge of the black hole are inferred 
from the asymptotic behavior of the metric and the dilaton at
spatial infinity.

\begin{figure}[t]
\begin{center}
\includegraphics[type=pdf,ext=.pdf,read=.pdf,width=8.0cm]{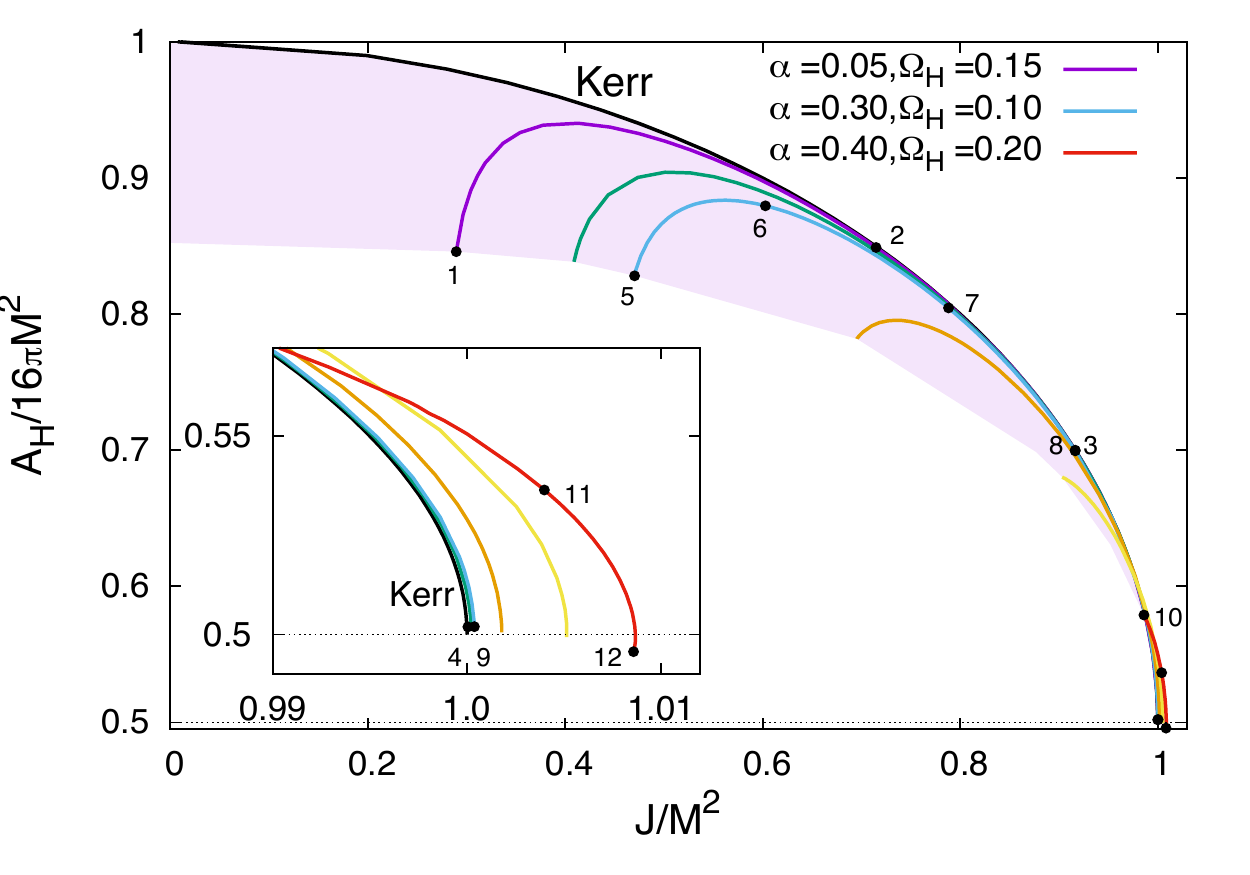}
\end{center}
\vspace{-0.5cm}
\caption{\textcolor{black}{
The scaled horizon area $a_{\rm H}=A_{\rm H}/16 \pi M^2$
is shown vs the scaled angular momentum $J/M^2$ for solutions with 
coupling constant $\gamma=1$.
The domain of existence is bounded by Kerr black holes
(black line),
and static (left boundary), 
critical (lower boundary to the left of Kerr)
and extremal (lower boundary to the right of Kerr, see inset) 
EdGB black holes.
The locations of the 12~configurations studied in this paper
are marked by the numbered black dots. The colored lines refer
to families of configurations with fixed product 
$\Omega_{\rm H} \alpha^{1/2}$.
} \label{f-domain}}
\end{figure}

In the present paper we study 12~numerical metrics for the dilaton
coupling $\gamma=1$, of which the values of the coupling $\alpha$, 
the scaled angular momentum $J/M^2$, and the scaled horizon area $a_{\rm H}$
are reported in Tab.~\ref{tab}. The locations of these 12~configurations 
in the domain of existence of the black hole solutions in EdGB gravity 
are shown in Fig.~\ref{f-domain}.

\begin{table*}[t]
\begin{tabular}{|c|ccc|c|c|}
\hline
\hspace{0.2cm} Solution \hspace{0.2cm} & \hspace{0.3cm} $\alpha$ \hspace{0.3cm}
 &\hspace{0.4cm} $J/M^2$ \hspace{0.4cm} & \hspace{0.5cm} $a_{\rm H}$ \hspace{0.5cm} & \hspace{0.3cm} $\chi^2_{\rm min,red}$ (NuSTAR) \hspace{0.3cm} & \hspace{0.1cm} $\chi^2_{\rm min,red}$ (LAD/eXTP) \hspace{0.1cm} \\
\hline 
1 & 0.05 & 0.289791  &  0.845924 & 1.1 & 1.9 \\
2 & 0.05 & 0.714817  &  0.848867 & 1.1 & 1.8 \\
3 & 0.05 & 0.916513  &  0.699935 & 0.9 & 2.5 \\
4 & 0.05 & 1.00005   &  0.501936 & 0.9 & 2.6 \\
\hline
5 & 0.3 & 0.470232   & 0.828229 & 1.0 & 1.8 \\
6 & 0.3 & 0.602776   & 0.879576 & 0.9 & 1.9 \\
7 & 0.3 & 0.788243   & 0.804400 & 1.1 & 2.7 \\
8 & 0.3 & 0.916448   & 0.699469 & 1.0 & 2.4 \\
9 & 0.3 & 1.00038    & 0.502024 & 1.0 & 2.2 \\
\hline
10 & 0.4 & 0.986010  &  0.578885 & 1.0 & 2.4 \\
11 & 0.4 & 1.00399   &  0.536395 & 1.0 & 2.2 \\
12 & 0.4 & 1.00858   &  0.495732 & 1.1 & 2.4 \\
\hline
\end{tabular}
\vspace{0.2cm}
\caption{The 12~numerical metrics studied in our analysis. For every metric, the values of its parameters $\alpha$, $J/M^2$, and $a_{\rm H}$ are shown in the second, third, and fourth columns, respectively. The fifth column shows the reduced $\chi^2$ of the best-fit in the NuSTAR simulations. The sixth column is for the reduced $\chi^2$ of the best-fit in the LAD/eXTP simulations. See the text for more details. \label{tab}}
\end{table*}

\section{X-ray reflection spectrum \label{s-iron}}

Within the disk-corona model~\cite{corona1,corona2}, an accreting black hole is surrounded by a geometrically thin and optically thick disk. The disk is in the equatorial plane, perpendicular to the black hole spin. In the Novikov-Thorne model~\cite{nt}, the disk emits like a blackbody locally and as a multi-color blackbody when integrated radially. The inner edge of the disk is at the innermost stable circular orbit (ISCO). The particles in the disk follow nearly-geodesic circular orbits in the equatorial plane. When they reach the ISCO radius, they quickly plunge onto the central object, so that the emission inside the ISCO can be usually ignored. The corona is a hotter ($\sim 100$~keV), usually optically thin, cloud around the black hole, but its exact geometry is currently unknown. For instance, it may be the base of the jet, an atmosphere just above the accretion disk, or the accretion flow between the inner edge of the disk and the black hole.

Because of inverse Compton scattering of the thermal photons from the accretion disk off the free electrons in the corona, the latter becomes a source of hard X-ray with a power-law spectrum $E^{-\Gamma}$. The photons of the corona can also illuminate the disk, producing a reflection component with some fluorescent emission lines~\cite{j}. The most prominent feature in the reflection spectrum is usually the iron K$\alpha$ line, which is at 6.4~keV in the case of neutral or weakly ionized iron and shifts up to 6.97~keV in the case of H-like iron ions.

The iron K$\alpha$ line is a very narrow feature in the rest-frame of the emitter. On the contrary, the line observed in the reflection spectrum of astrophysical black holes is broad and skewed, as a result of special and general relativistic effects (gravitational redshift, Doppler boosting, light bending) occurring in the strong gravity region of the black hole. In the presence of high-quality data and with the correct astrophysical model, the analysis of the iron line can be a powerful tool to probe the near horizon region. This technique was proposed and developed to estimate the black hole spin under the assumption of the Kerr background~\cite{iron1,iron2}, and only more recently it has been extended to test alternative theories of gravity~\cite{i1,i2,i3,i4}. The technique is often called the iron line method, because the iron K$\alpha$ line is the most prominent feature, but any measurement of the spacetime metric around black holes should be done by fitting the whole reflection spectrum, not only the iron line.

The shape of the iron line as detected in the flat faraway region is determined by the spacetime metric, the inclination angle of the disk with respect to the line of sight of the distant observer, and the geometry and the intensity profile of the emitting region. The disk is usually assumed completely axisymmetric, and emitting from the ISCO radius to some large radius. The intensity profile is actually a crucial ingredient and depends on the exact geometry of the corona, which, unfortunately, is currently unknown. The intensity profile for a corona with arbitrary geometry is often approximated by a power-law ($\propto 1/r^q$, where $q$ is the emissivity index) or by a broken power-law ($\propto 1/r^{q_1}$ for $r < r_{\rm br}$ and $\propto 1/r^{q_2}$ for $r > r_{\rm br}$, where $q_1$ and $q_2$ are, respectively, the inner and the outer emissivity indices and $r_{\rm br}$ is the breaking radius).

\begin{figure*}[t]
\begin{center}
\includegraphics[type=pdf,ext=.pdf,read=.pdf,width=9cm]{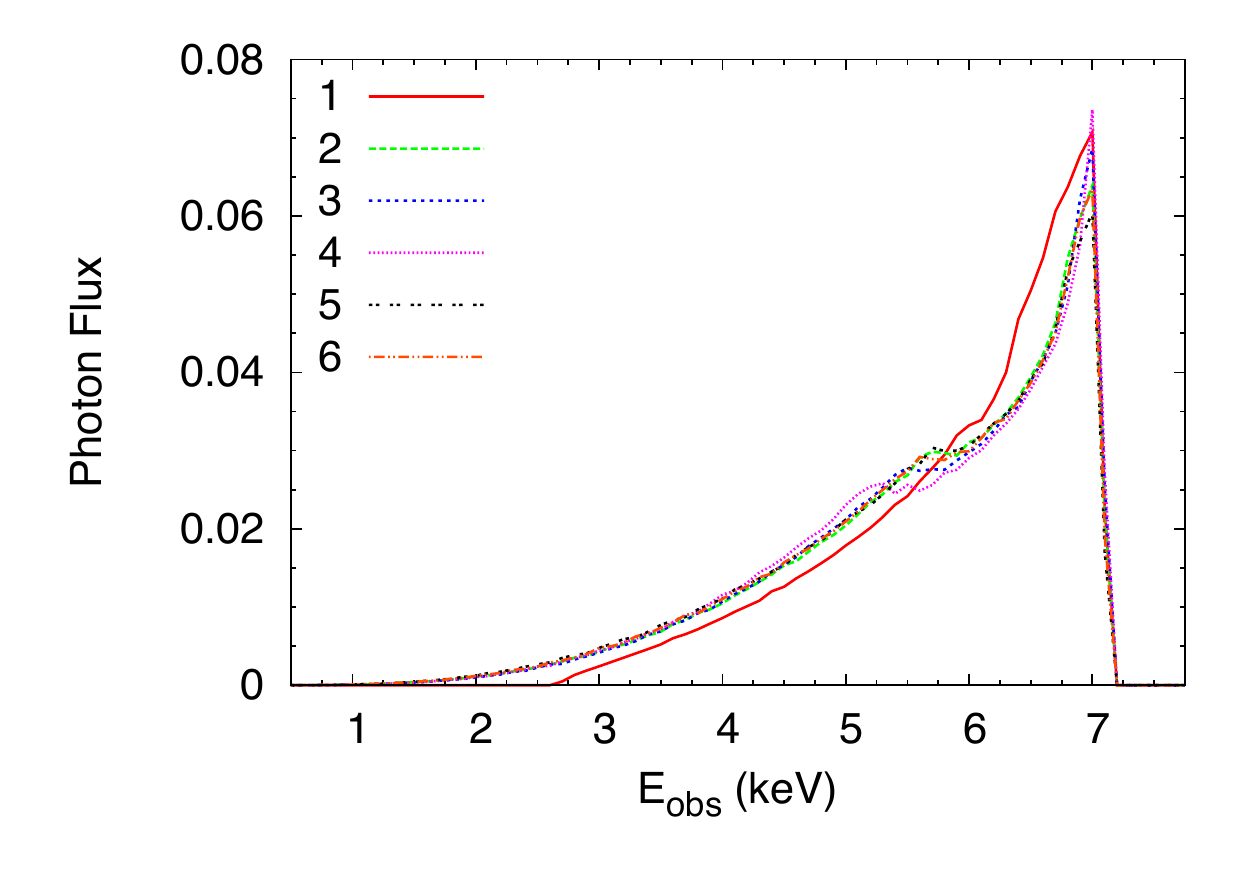}
\hspace{-0.5cm}
\includegraphics[type=pdf,ext=.pdf,read=.pdf,width=9cm]{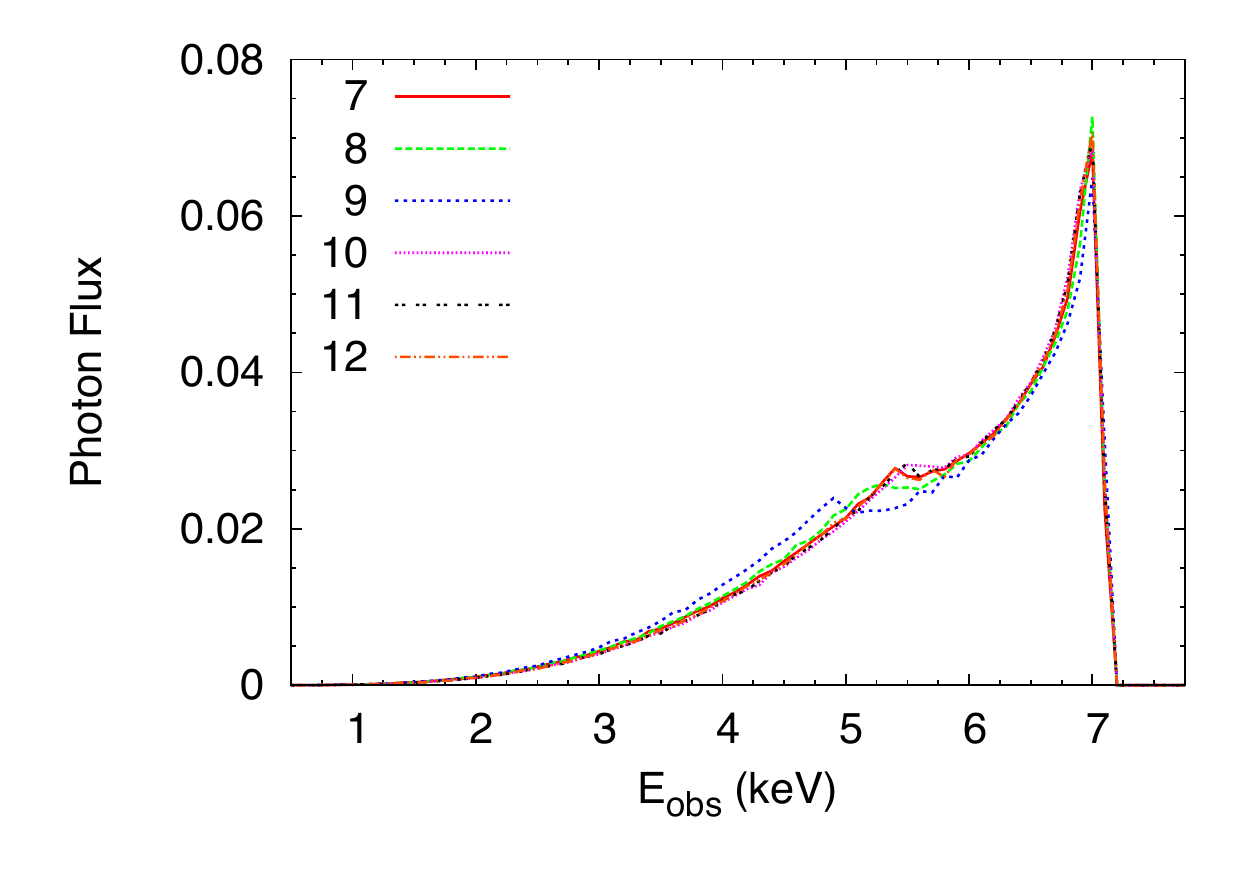}
\end{center}
\vspace{-0.7cm}
\caption{Iron line shapes for solutions~1-6 (left panel) and 7-12 (right panel). The intensity profile is $1/r^3$, the viewing angle is $i = 45^\circ$, and the energy of the line in the rest frame of the emitting gas is at 6.4~keV. \label{f-iron}}
\end{figure*}

Fig.~\ref{f-iron} shows the iron line shapes calculated in black hole solutions~1-12, assuming that the intensity profile scales as $1/r^3$ (Newtonian limit at large radii for a lamppost corona), that the inclination angle of the disk with respect to the line of sight of the distant observer is $i=45^\circ$, and that the rest frame energy of the line is 6.4~keV. The calculations are done with the code described in~\cite{cfm2,i3} and extended to treat numerical metrics in Ref.~\cite{x1}.

\section{Simulations \label{s-sim}}

As a preliminary analysis to figure out whether present and future X-ray missions can distinguish the Kerr black holes of Einstein's gravity from the black holes in EdGB gravity, we follow the approach already employed in Refs.~\cite{x1,x2,x3,x4,x5,x6} to study the possibility of testing a number of non-Kerr metrics. We simulate an observation with a specific instrument employing the iron line calculated in the non-Kerr metric, and we fit the simulated data with the iron line of a Kerr model. If the latter can provide a good fit, we can conclude that X-ray reflection spectroscopy cannot distinguish that black hole from those in Einstein's gravity. If it is not possible to get a good fit, the model can be tested. Note that current observations can be fitted with a Kerr model. This means that we could rule out some spacetimes if we find that simulations with current X-ray missions cannot be fitted with a Kerr model.

We simulate observations with NuSTAR\footnote{http://www.nustar.caltech.edu} and LAD/eXTP\footnote{http://www.isdc.unige.ch/extp/}~\cite{extp}. The former is used to study the detection possibilities with current X-ray missions, the latter to explore the opportunities offered by the next generation of facilities. We do not consider a specific source, but we employ reasonable parameters for a bright black hole binary, which should be the kind of source most suitable for these tests. We model the spectrum of our source with a power-law (representing the primary component from the corona) and a single iron line (describing the reflection component). The energy flux of the source in the 1-10~keV range is $10^{-9}$~erg/s/cm$^2$ and the exposure time of the observation is 100~ks. We assume that the photon index of the power-law component is $\Gamma = 1.6$ and that the equivalent width of the iron line is 200~eV. We employ the iron lines shown in Fig.~\ref{f-iron}, where the viewing angle is $i = 45^\circ$ and the intensity profile scales as $1/r^3$.

The simulated observations are then treated as real data. After rebinning to ensure a minimum photon count per bin of 20 in order to use the $\chi^2$ statistics, we fit the data with a power-law and an iron line for Kerr spacetimes. For the iron line, we use RELLINE~\cite{relline}. There are 6~free parameters in the fit for the simulations with NuSTAR: the photon index of the power law $\Gamma$, the normalization of the power-law, the emissivity index $q$ for the intensity profile, the spin parameter $a_*$, the inclination angle of the disk $i$, and the normalization of the iron line. In the case of the simulations with LAD/eXTP, we have one more free parameter, the outer edge of the accretion disk, because the quality of the data is so good that it has its signature in the iron line and cannot be ignored. Note that the inner edge of the disk is set at the ISCO radius, so it only depends on the spacetime metric.

\begin{figure*}[t]
\begin{center}
\includegraphics[type=pdf,ext=.pdf,read=.pdf,width=9cm]{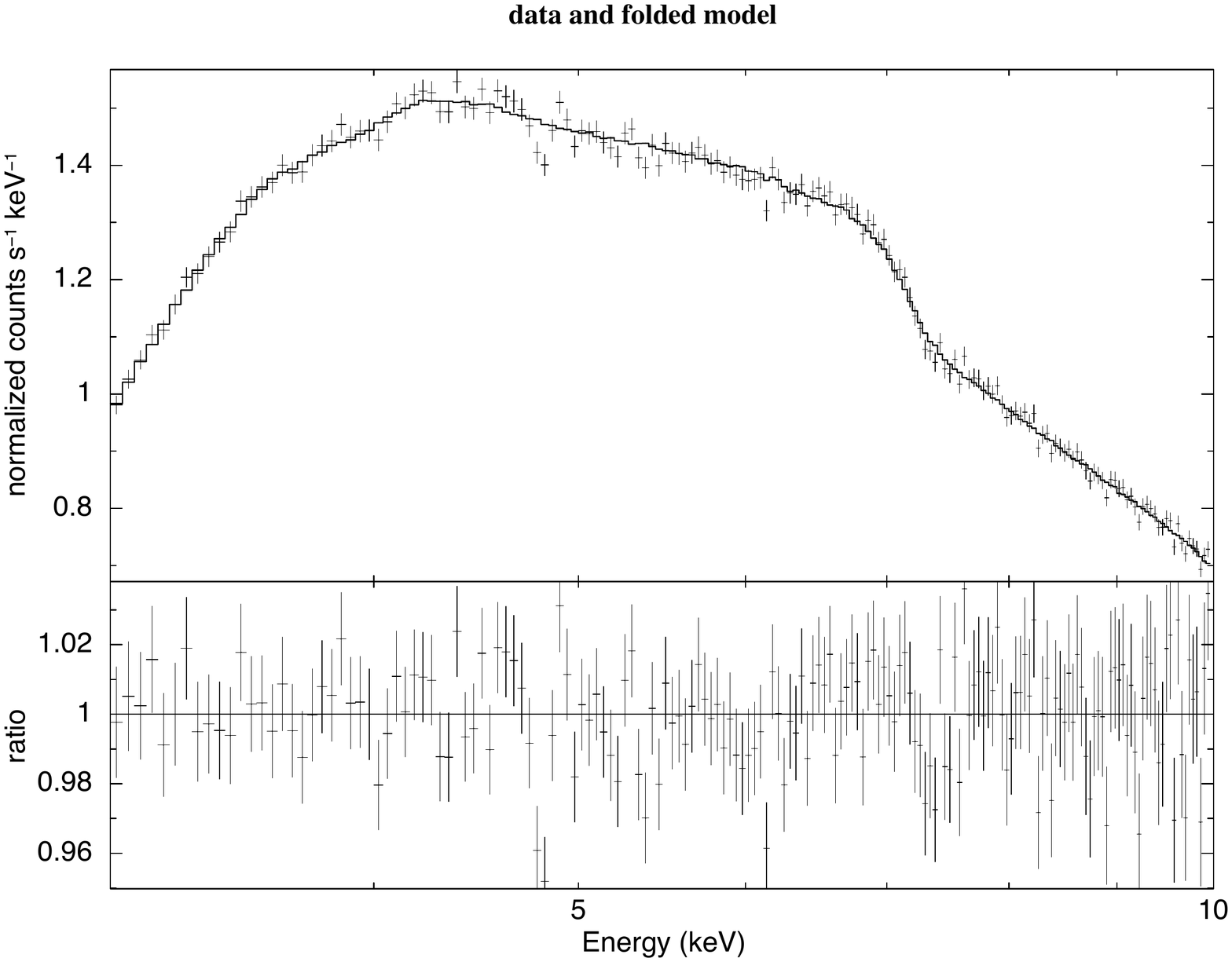}
\hspace{-0.5cm}
\includegraphics[type=pdf,ext=.pdf,read=.pdf,width=9cm]{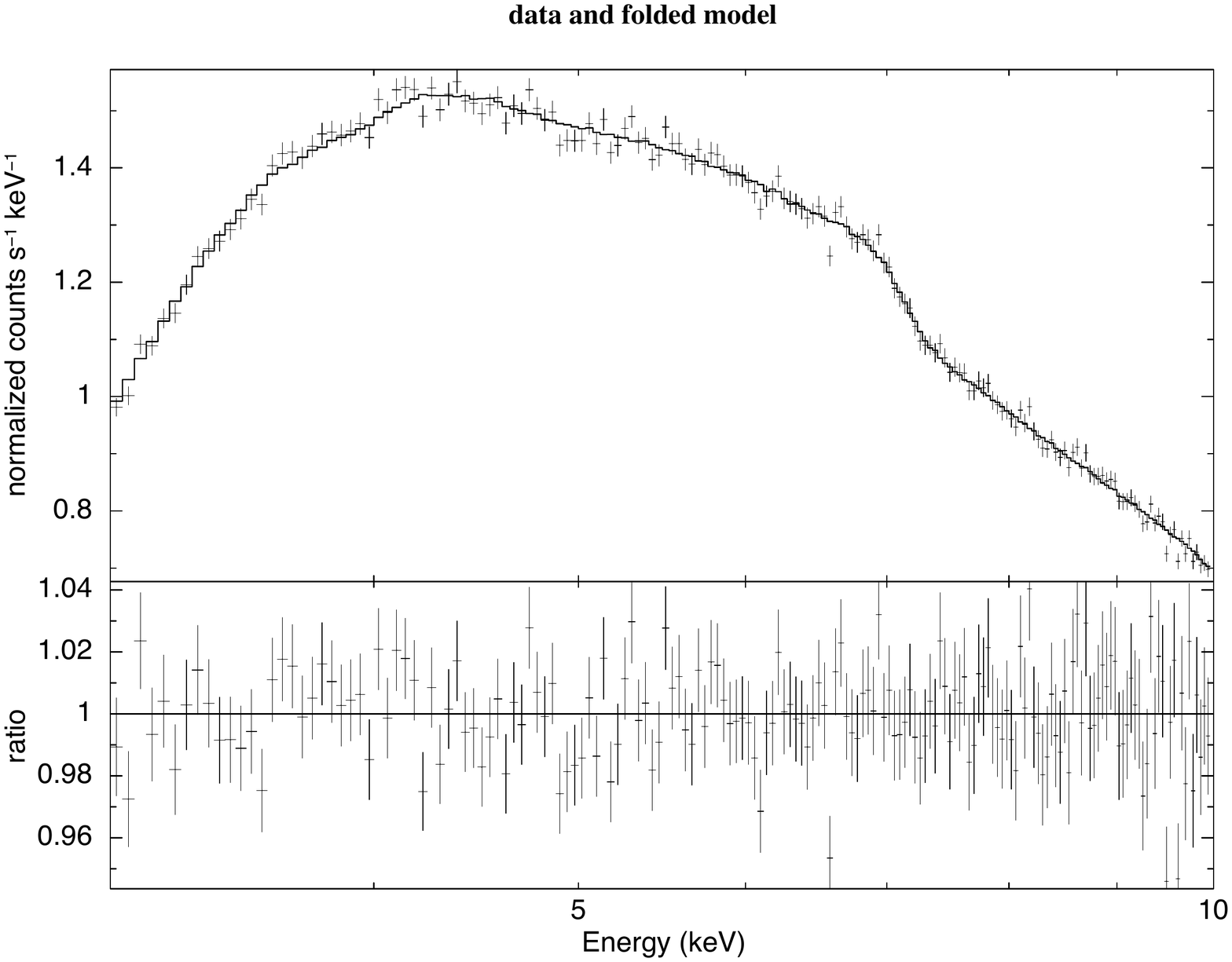} \\
\includegraphics[type=pdf,ext=.pdf,read=.pdf,width=9cm]{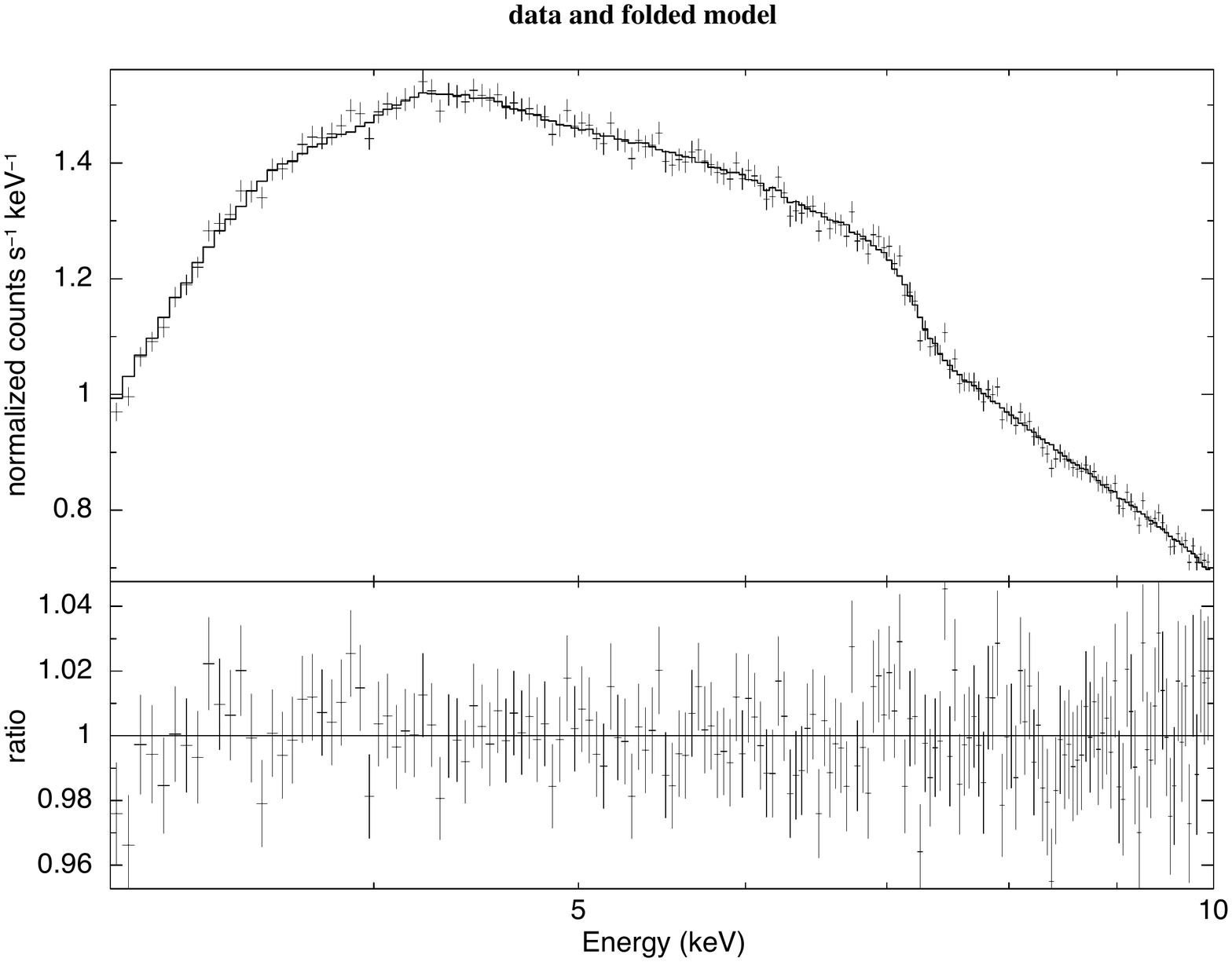}
\hspace{-0.5cm}
\includegraphics[type=pdf,ext=.pdf,read=.pdf,width=9cm]{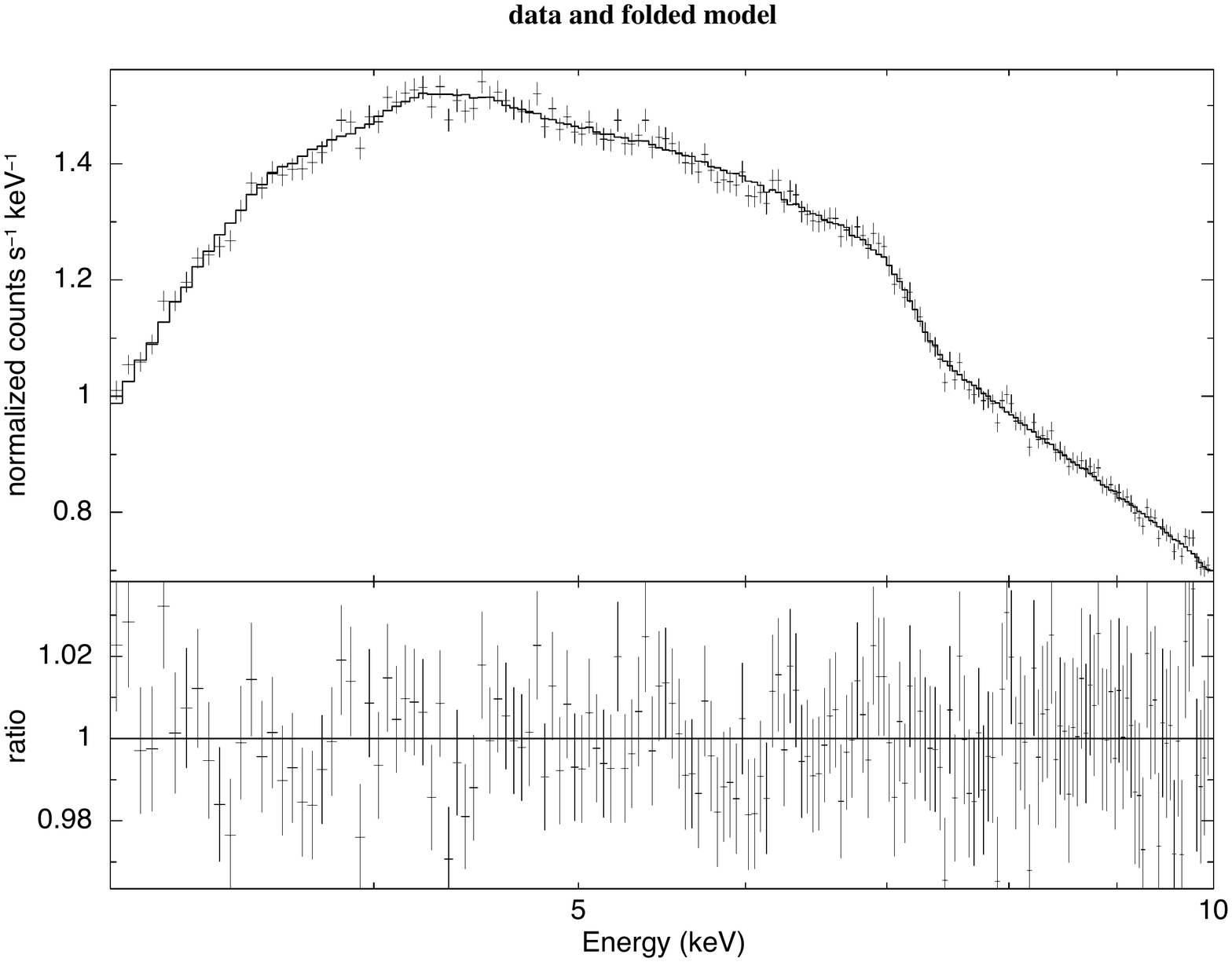} \\
\includegraphics[type=pdf,ext=.pdf,read=.pdf,width=9cm]{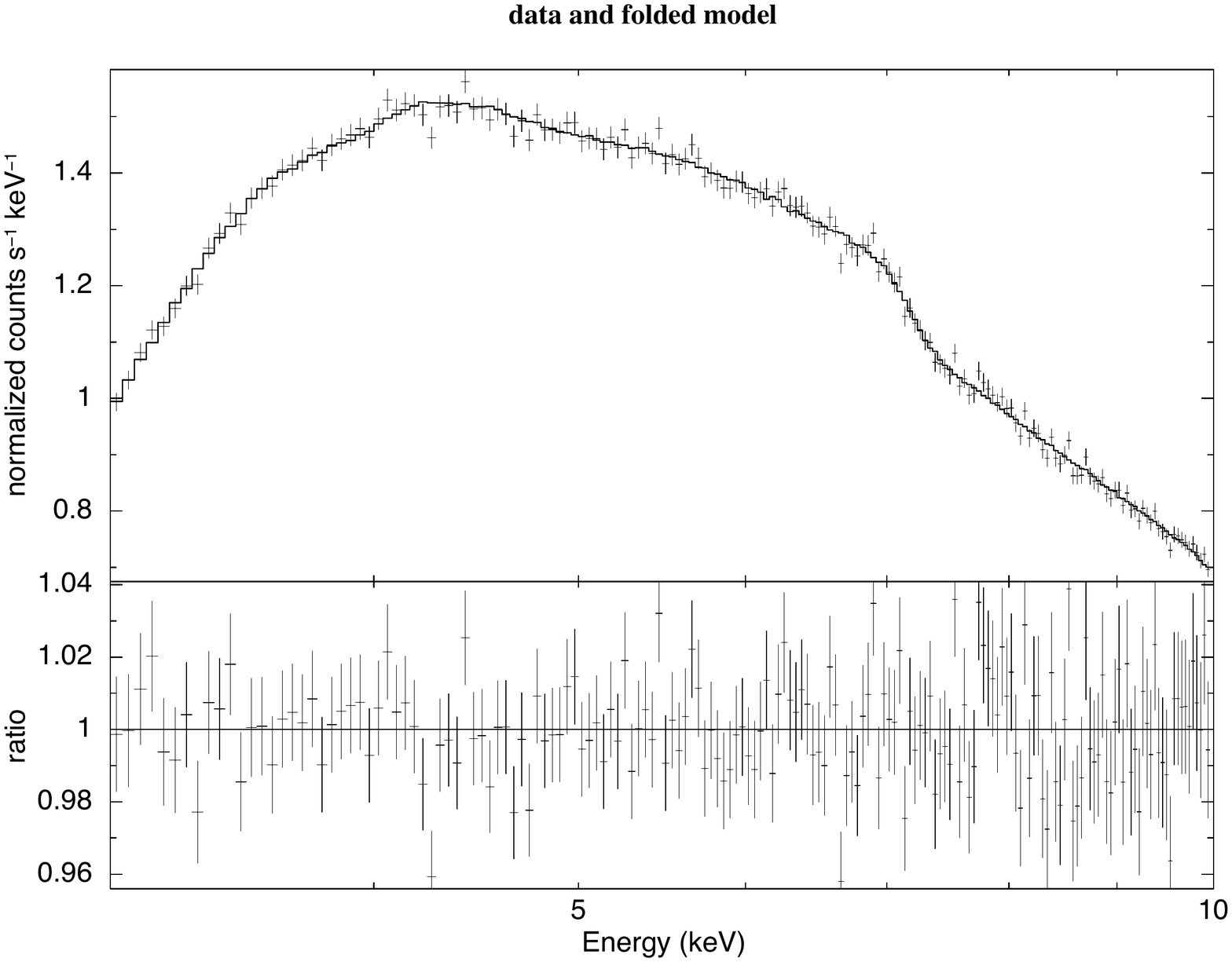}
\hspace{-0.5cm}
\includegraphics[type=pdf,ext=.pdf,read=.pdf,width=9cm]{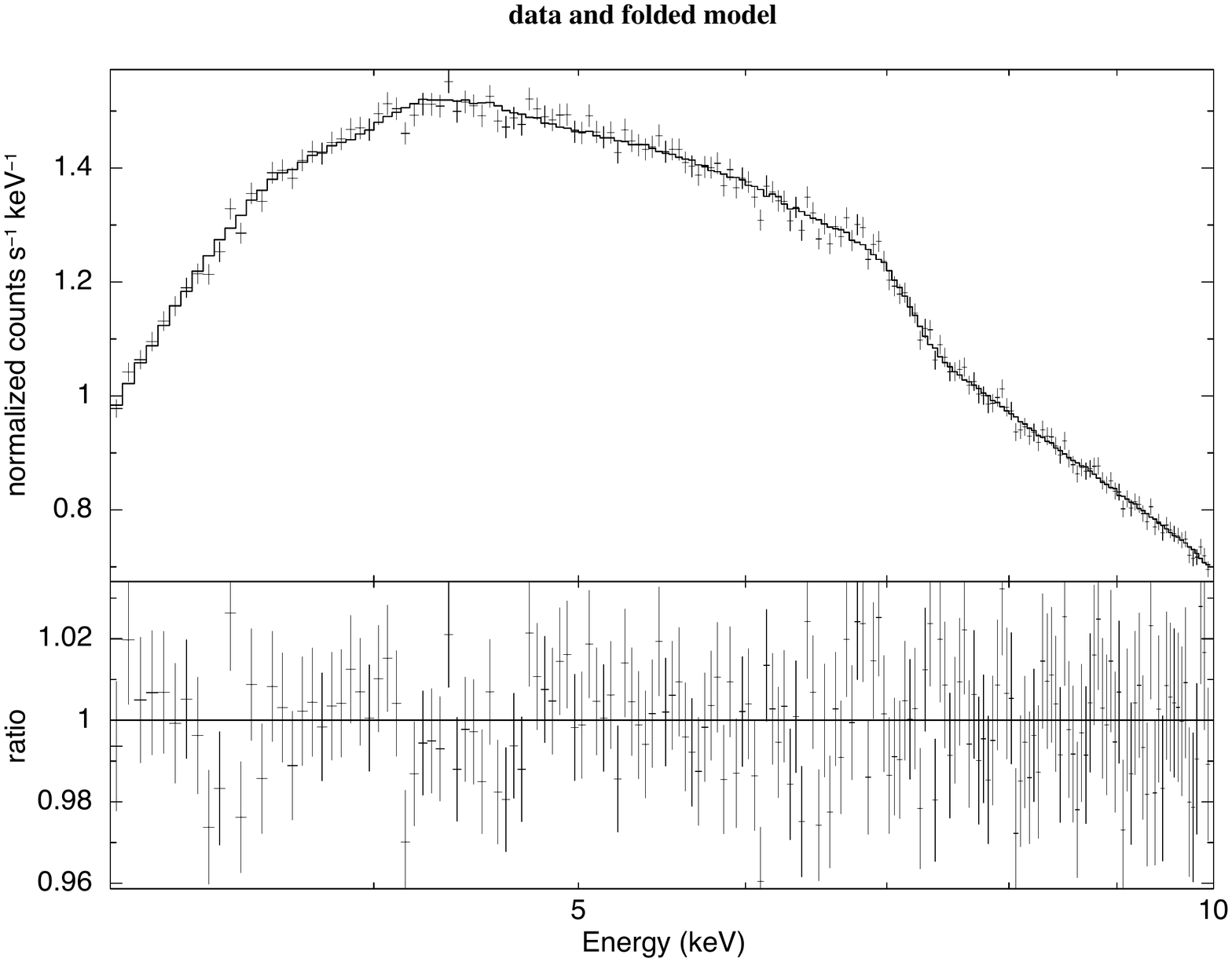}
\end{center}
\vspace{-0.5cm}
\caption{Results of our simulations with NuSTAR for 100~ks observations of a bright black hole binary. In each panel, the top quadrant shows the simulated data and the best-fit, while the bottom quadrant shows the ratio between the simulated data and the best-fit. The metric of the spacetime is described by Solution~1 (top left panel), Solution~2 (top right panel), Solution~3 (central left panel), Solution~4 (central right panel), Solution~5 (bottom left panel), and Solution~6 (bottom right panel). The inclination angle of the disk is $i = 45^\circ$ and the intensity profile is modeled with a power-law with emissivity index 3. See the text for more details. \label{f-nustar-a}}
\end{figure*}

\begin{figure*}[t]
\begin{center}
\includegraphics[type=pdf,ext=.pdf,read=.pdf,width=9cm]{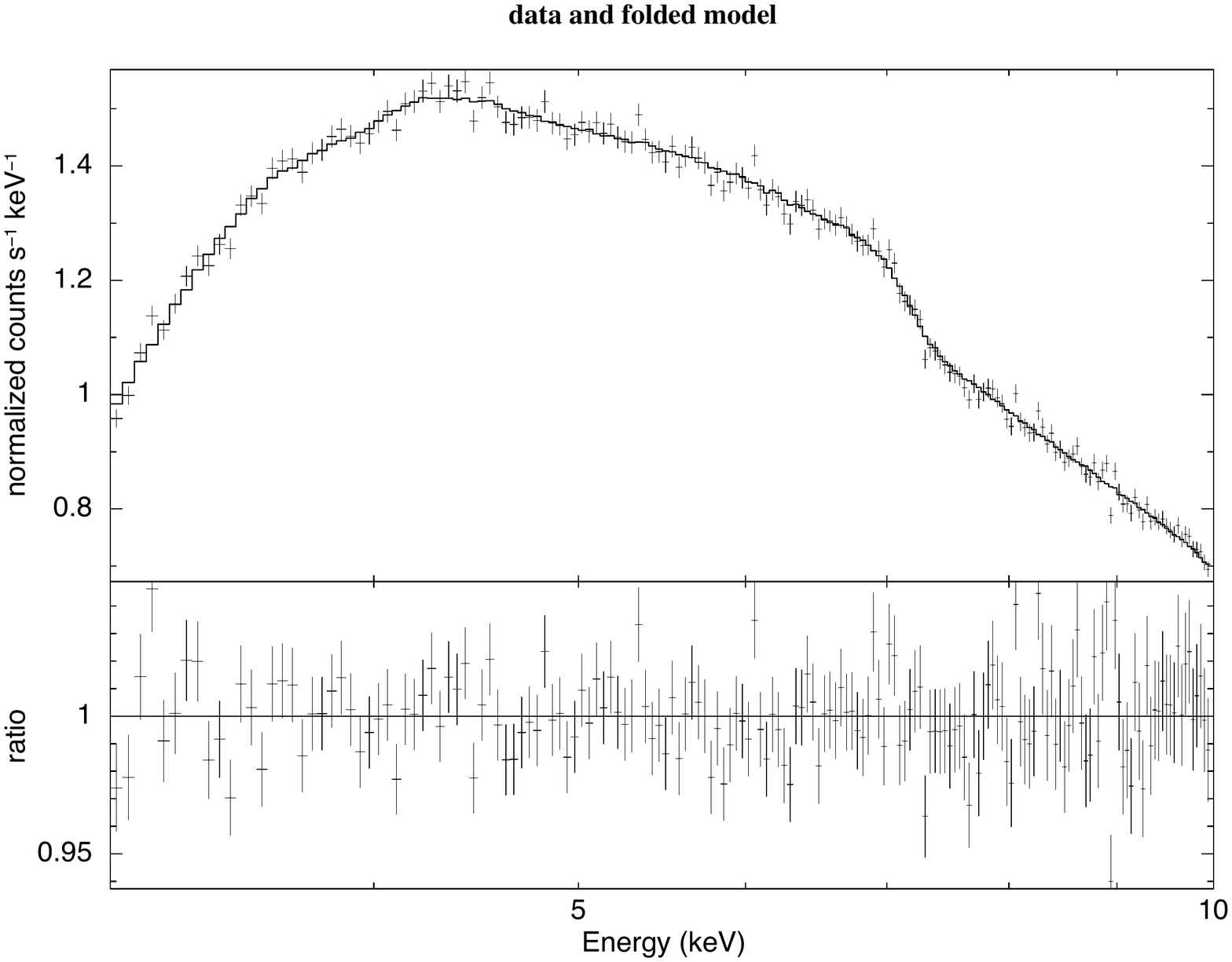}
\hspace{-0.5cm}
\includegraphics[type=pdf,ext=.pdf,read=.pdf,width=9cm]{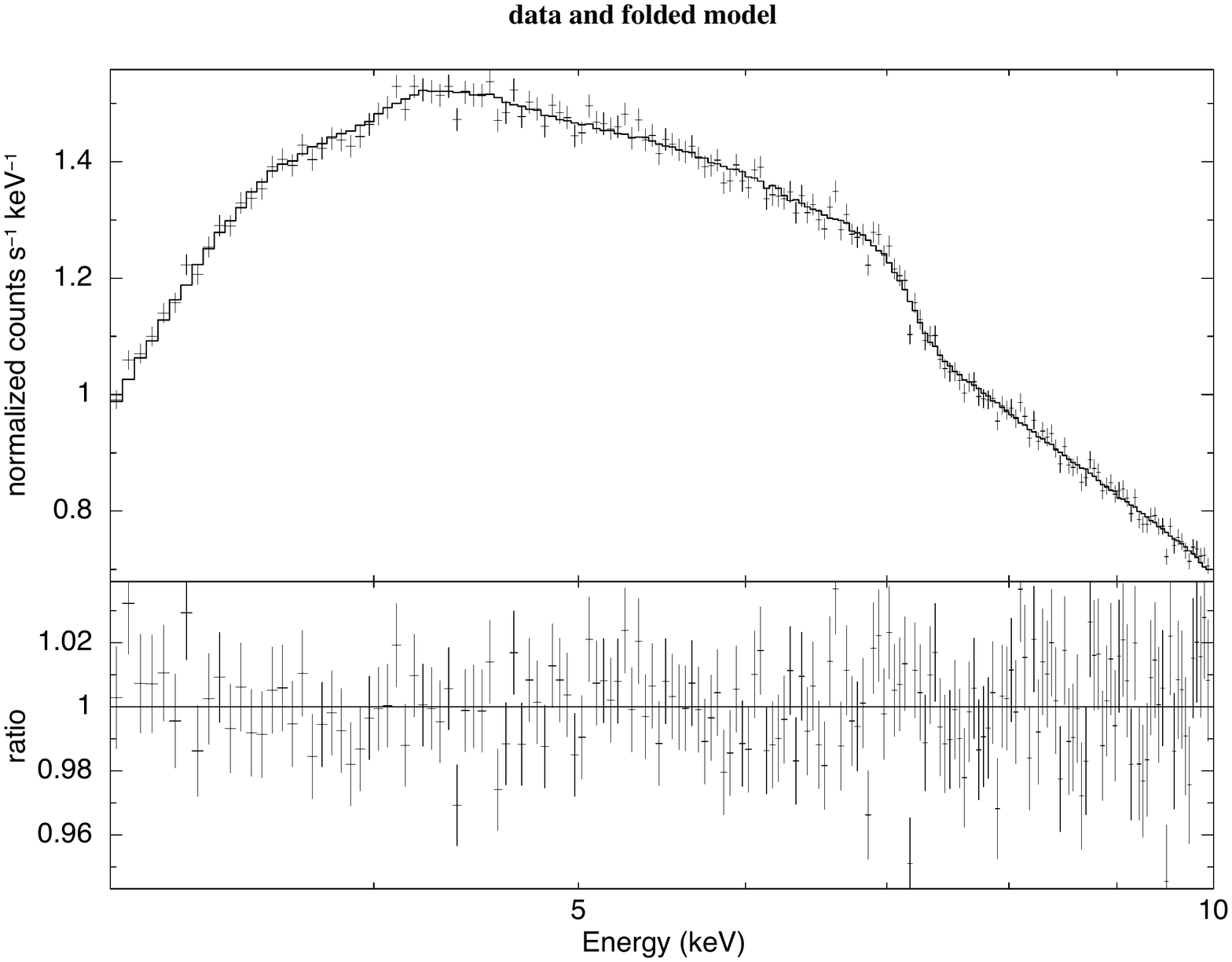} \\
\includegraphics[type=pdf,ext=.pdf,read=.pdf,width=9cm]{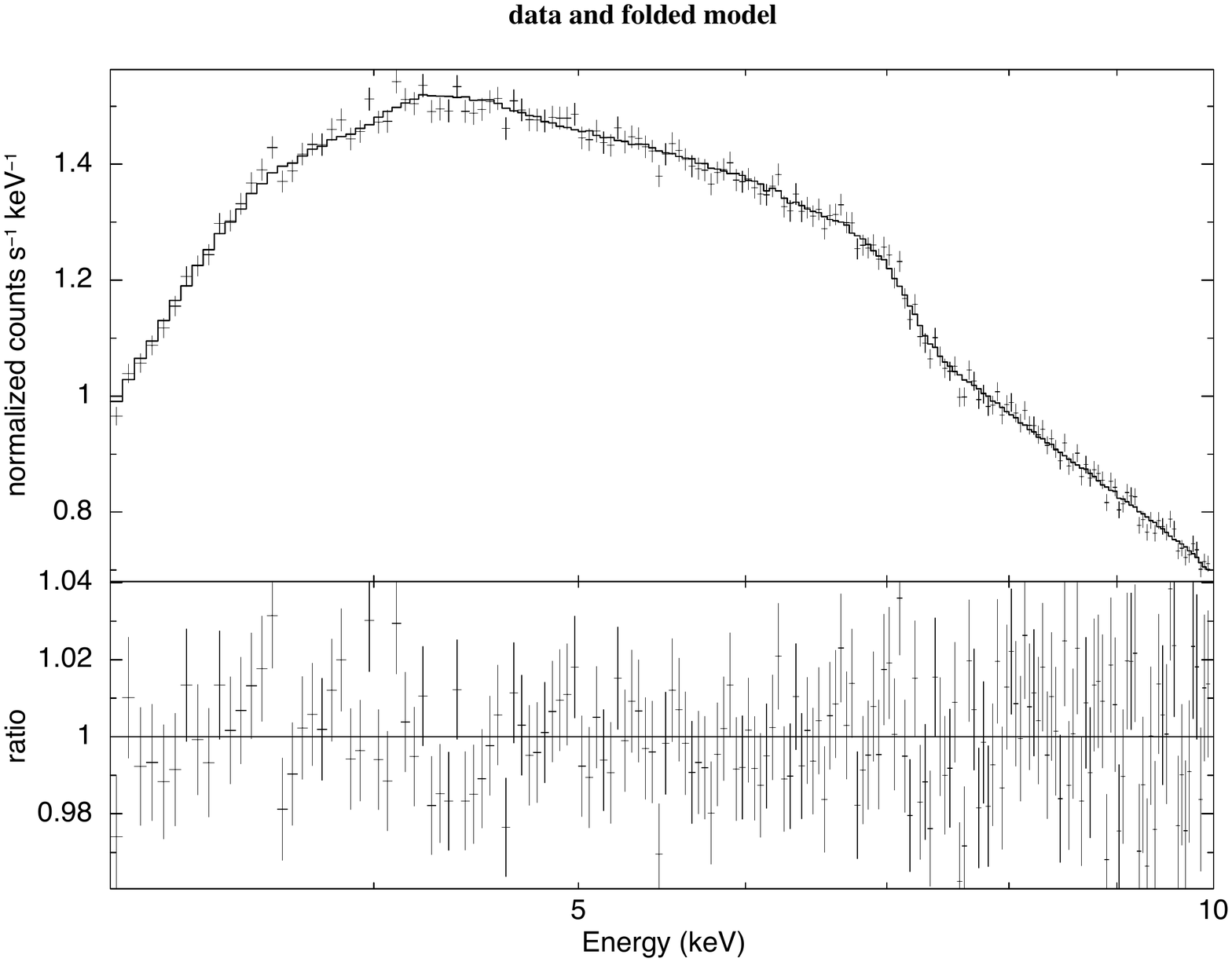}
\hspace{-0.5cm}
\includegraphics[type=pdf,ext=.pdf,read=.pdf,width=9cm]{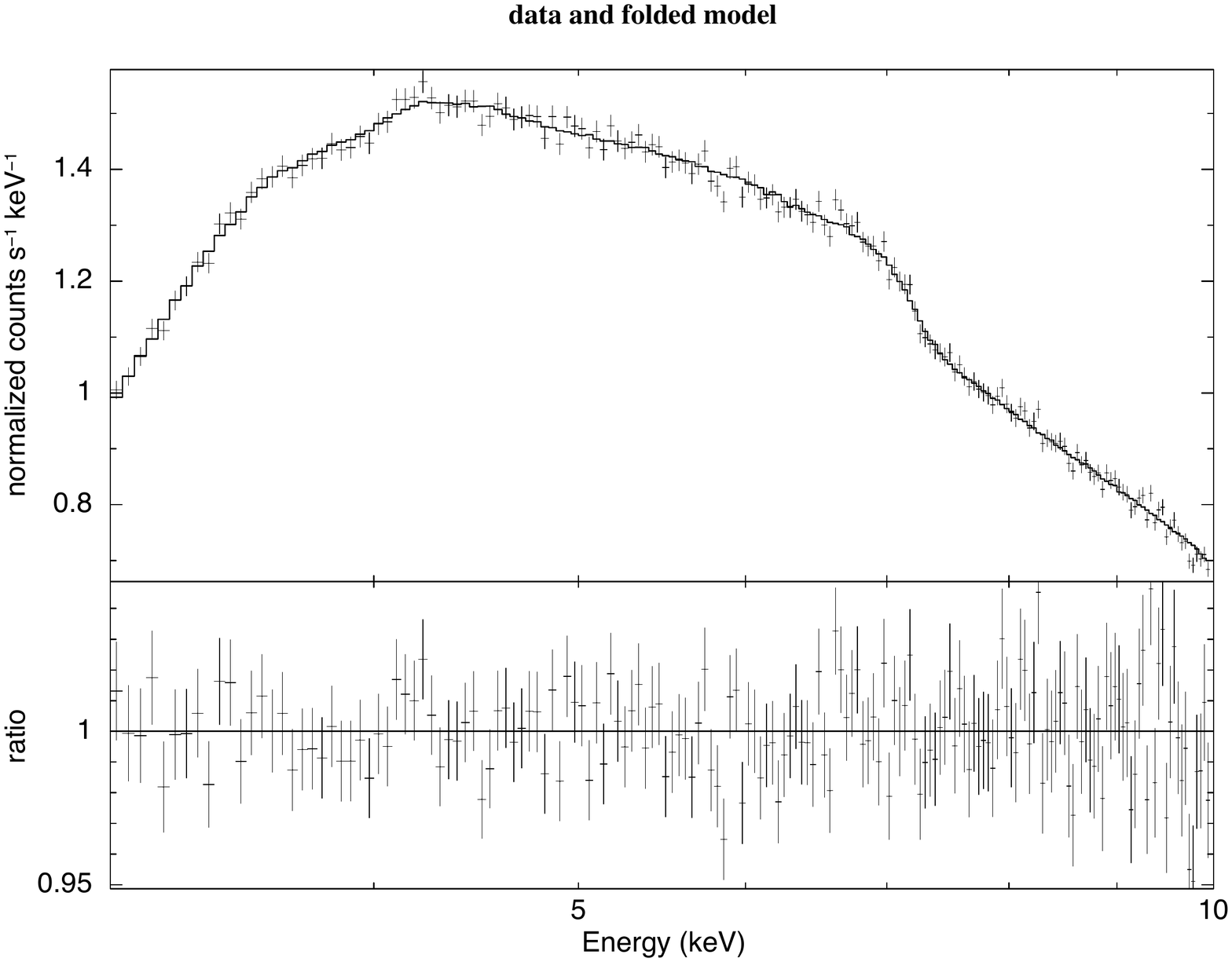} \\
\includegraphics[type=pdf,ext=.pdf,read=.pdf,width=9cm]{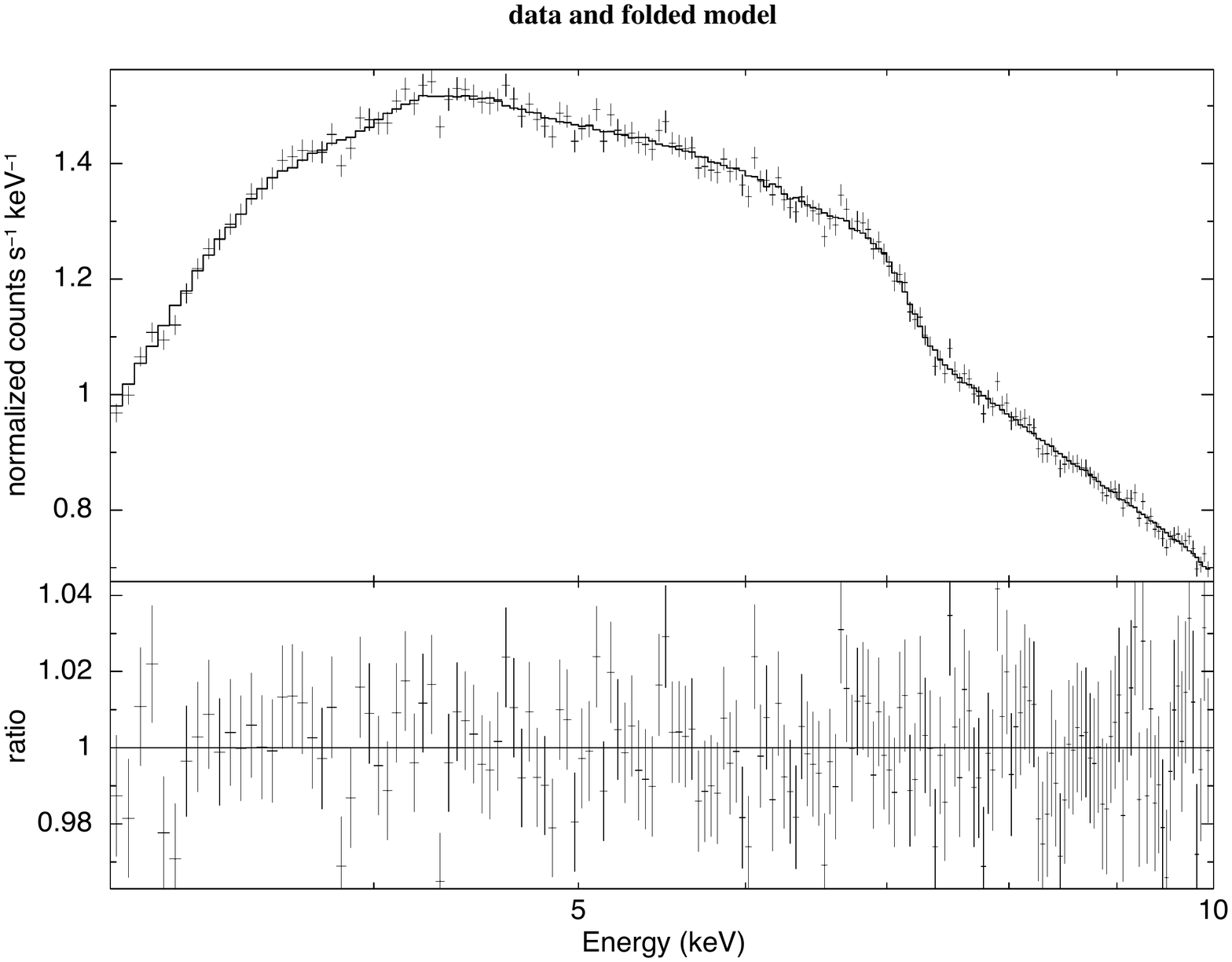}
\hspace{-0.5cm}
\includegraphics[type=pdf,ext=.pdf,read=.pdf,width=9cm]{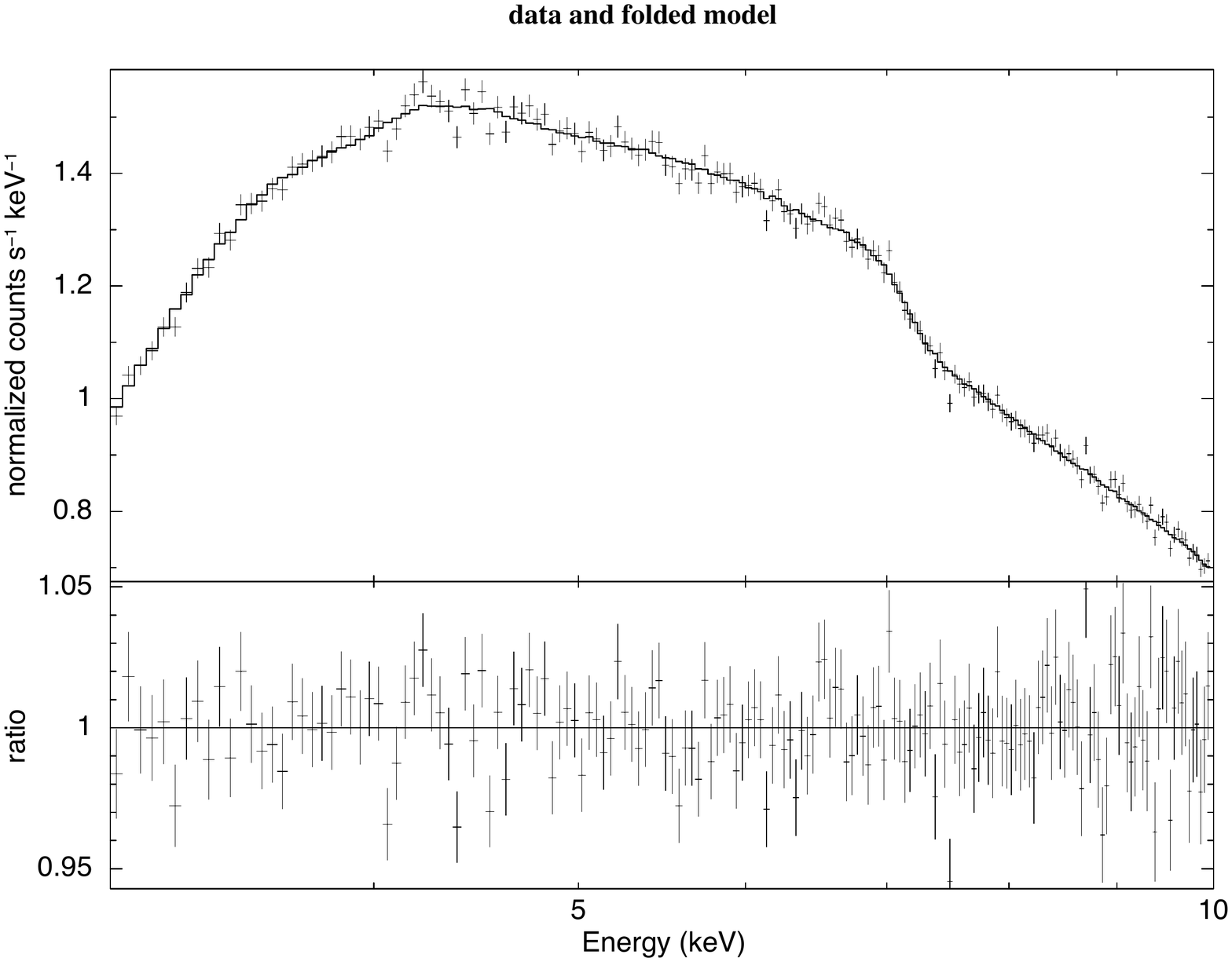}
\end{center}
\vspace{-0.5cm}
\caption{As in Fig.~\ref{f-nustar-a} for Solution~7 (top left panel), Solution~8 (top right panel), Solution~9 (central left panel), Solution~10 (central right panel), Solution~11 (bottom left panel), and Solution~12 (bottom right panel). \label{f-nustar-b}}
\end{figure*}

\begin{figure*}[t]
\begin{center}
\includegraphics[type=pdf,ext=.pdf,read=.pdf,width=9cm]{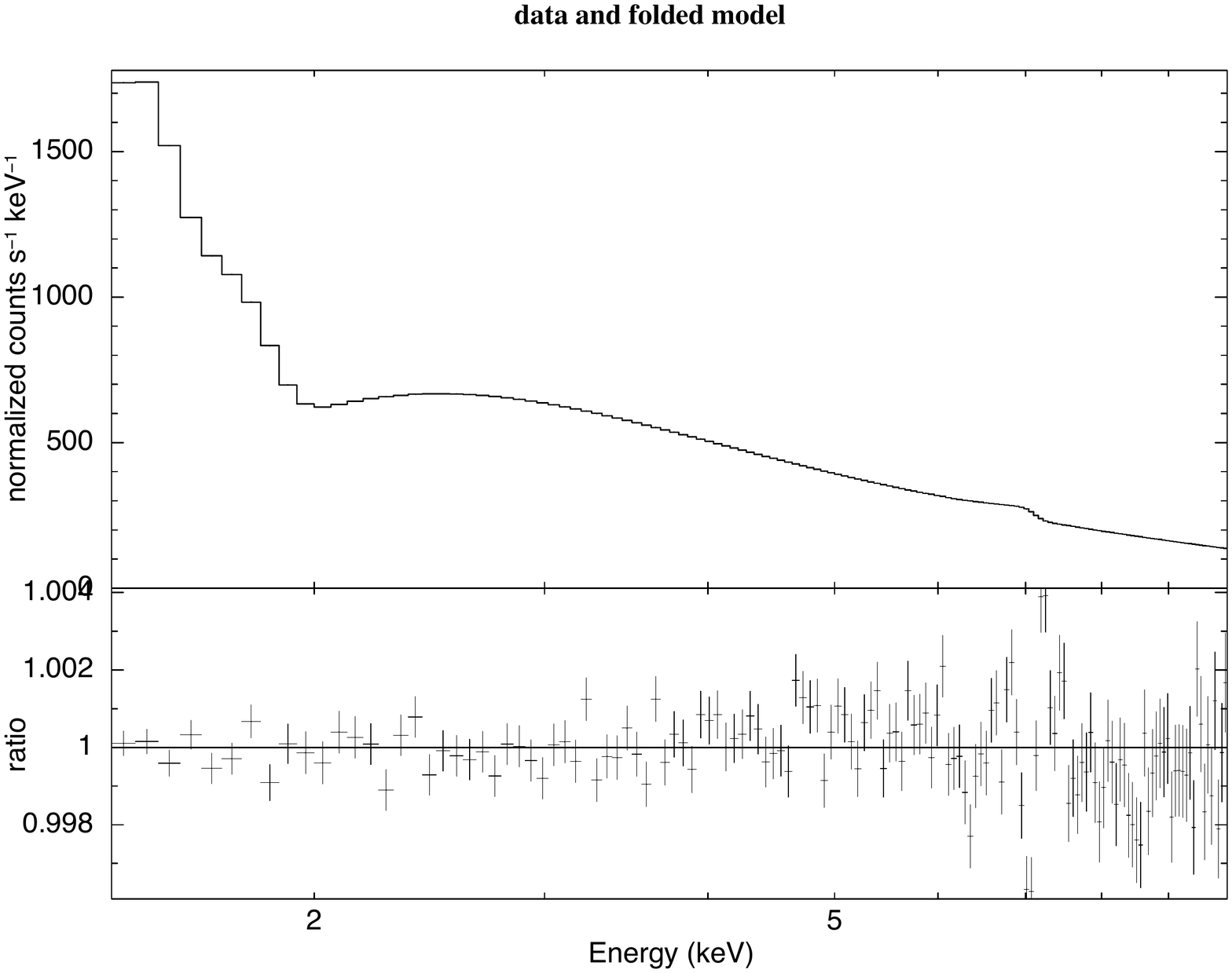}
\hspace{-0.5cm}
\includegraphics[type=pdf,ext=.pdf,read=.pdf,width=9cm]{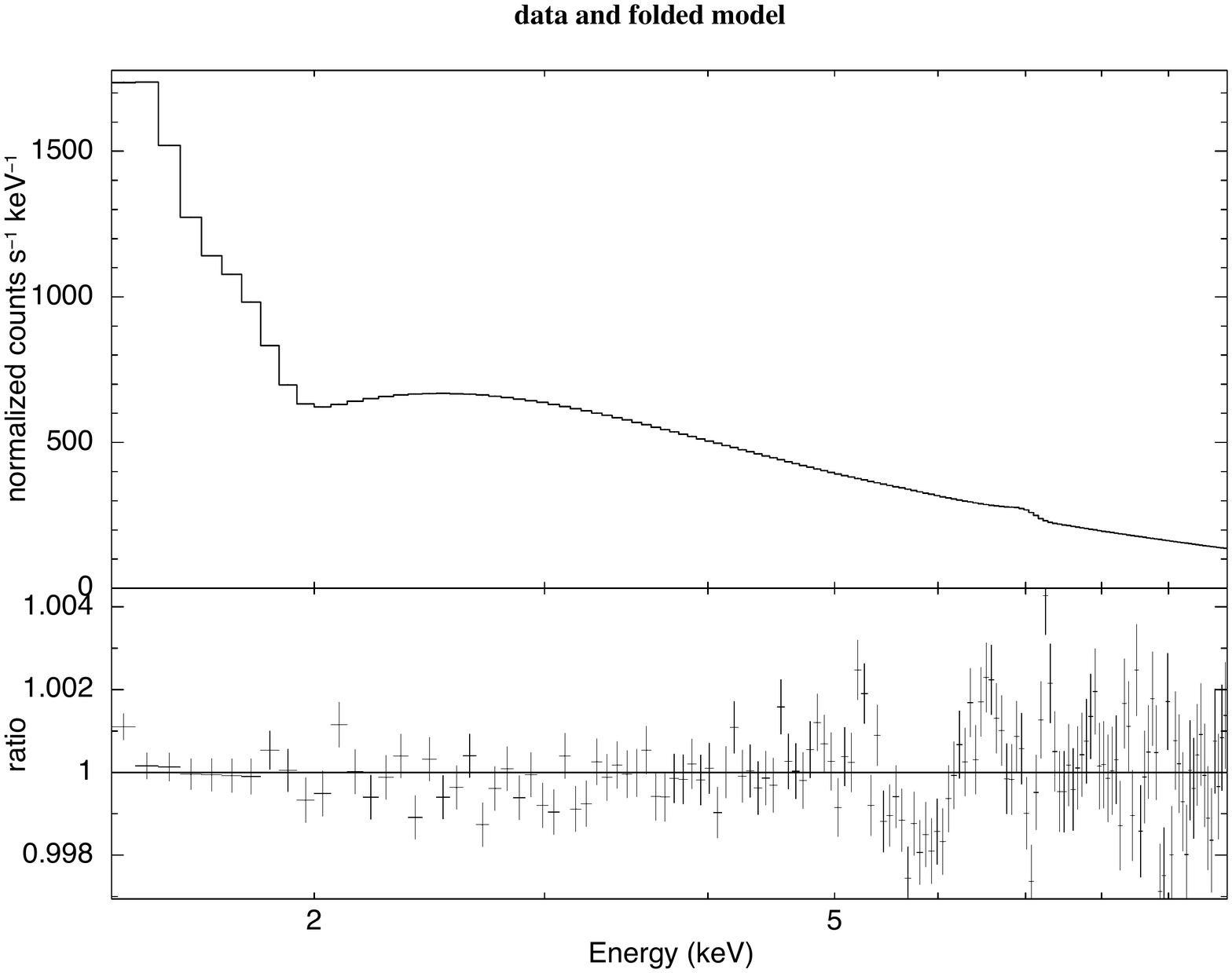} \\
\includegraphics[type=pdf,ext=.pdf,read=.pdf,width=9cm]{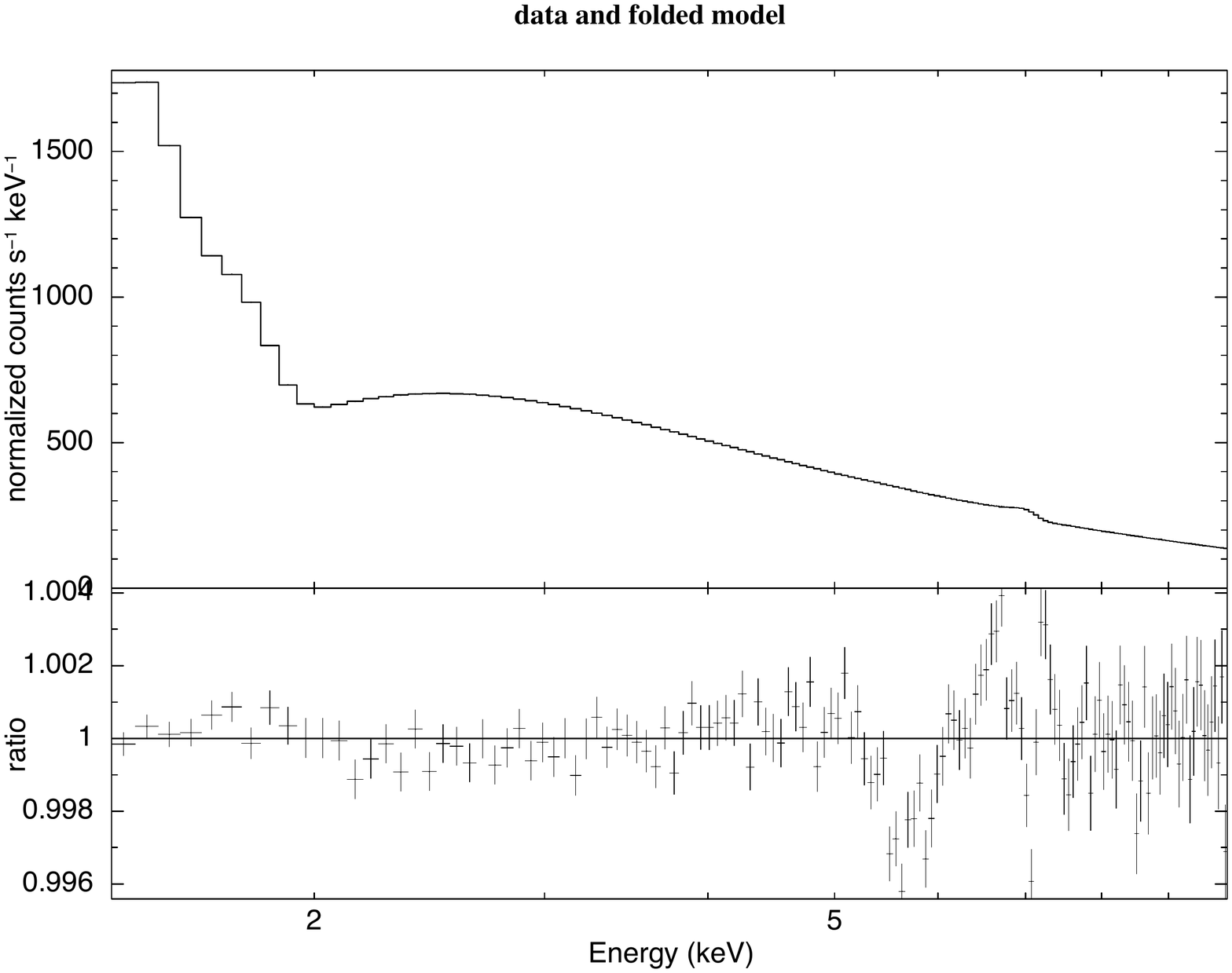}
\hspace{-0.5cm}
\includegraphics[type=pdf,ext=.pdf,read=.pdf,width=9cm]{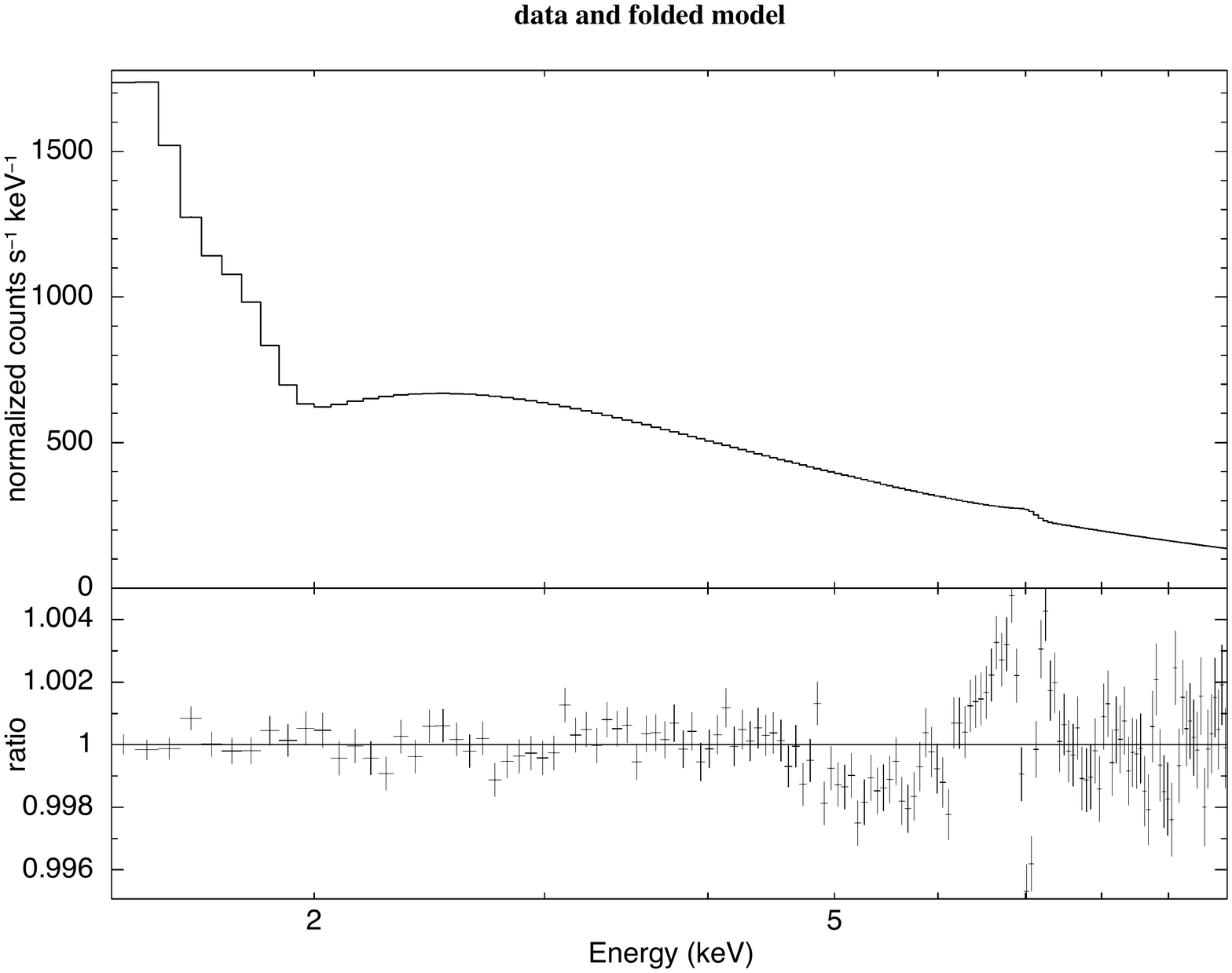} \\
\includegraphics[type=pdf,ext=.pdf,read=.pdf,width=9cm]{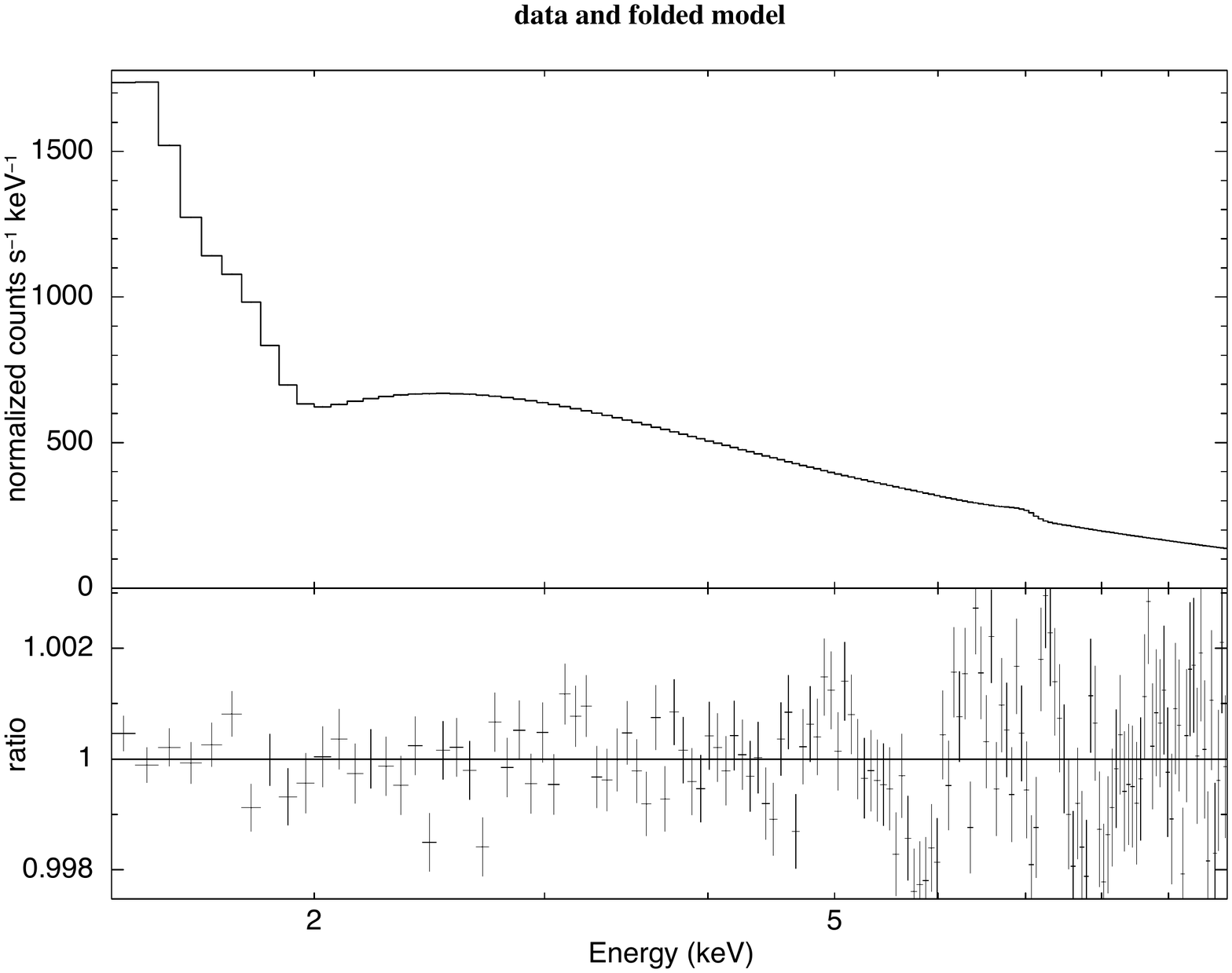}
\hspace{-0.5cm}
\includegraphics[type=pdf,ext=.pdf,read=.pdf,width=9cm]{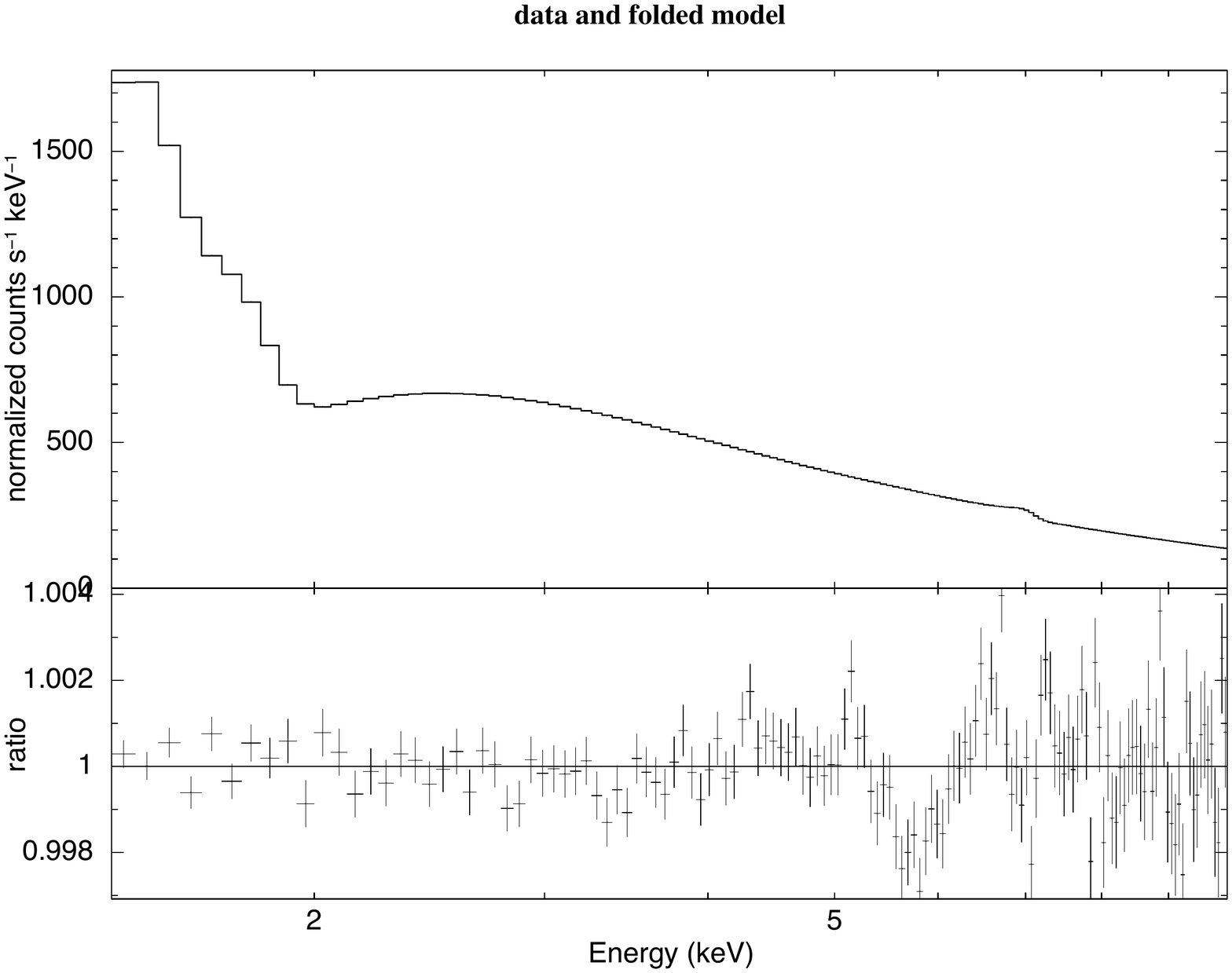}
\end{center}
\vspace{-0.5cm}
\caption{As in Fig.~\ref{f-nustar-a} for the simulations with LAD/eXTP. \label{f-lad-a}}
\end{figure*}

\begin{figure*}[t]
\begin{center}
\includegraphics[type=pdf,ext=.pdf,read=.pdf,width=9cm]{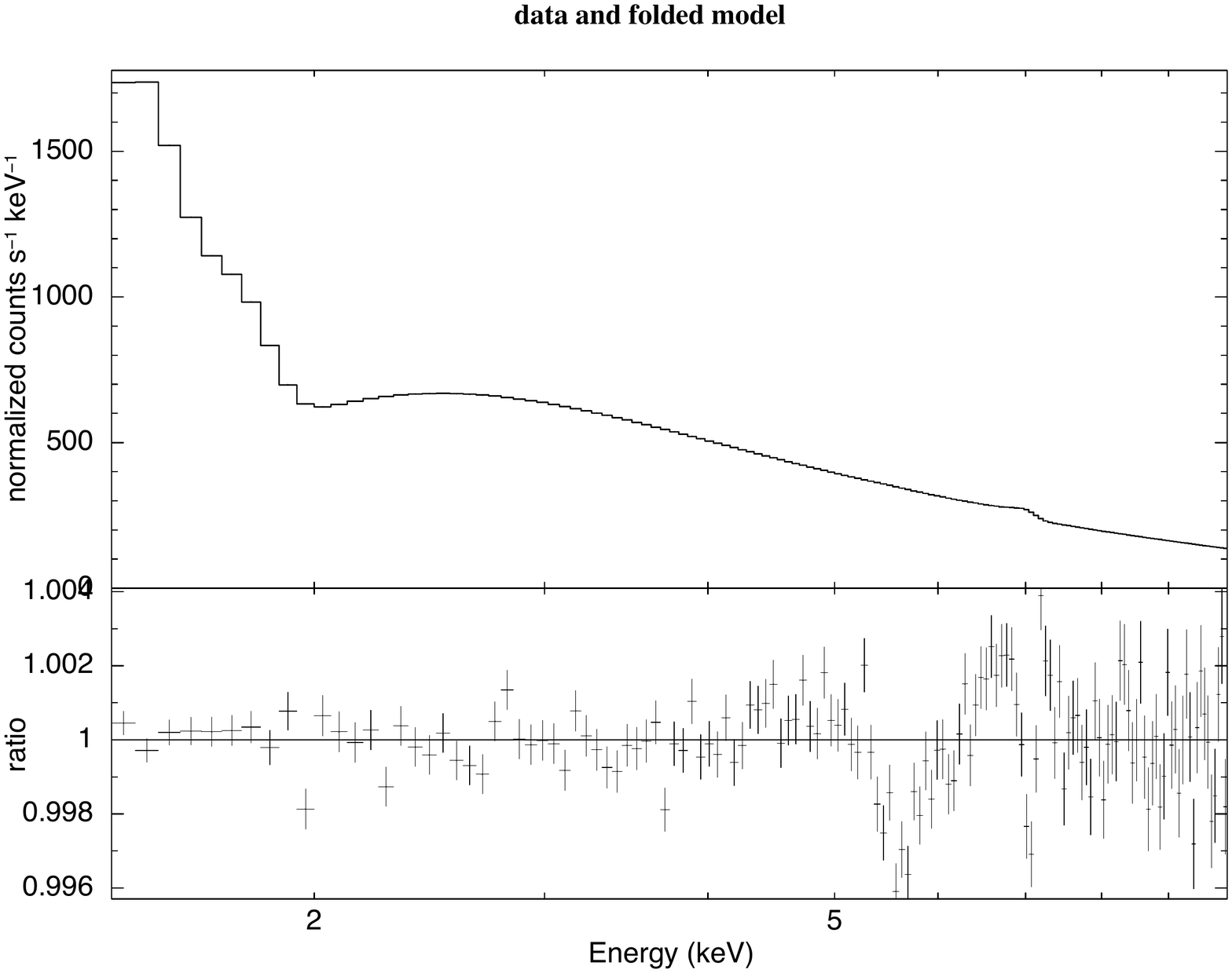}
\hspace{-0.5cm}
\includegraphics[type=pdf,ext=.pdf,read=.pdf,width=9cm]{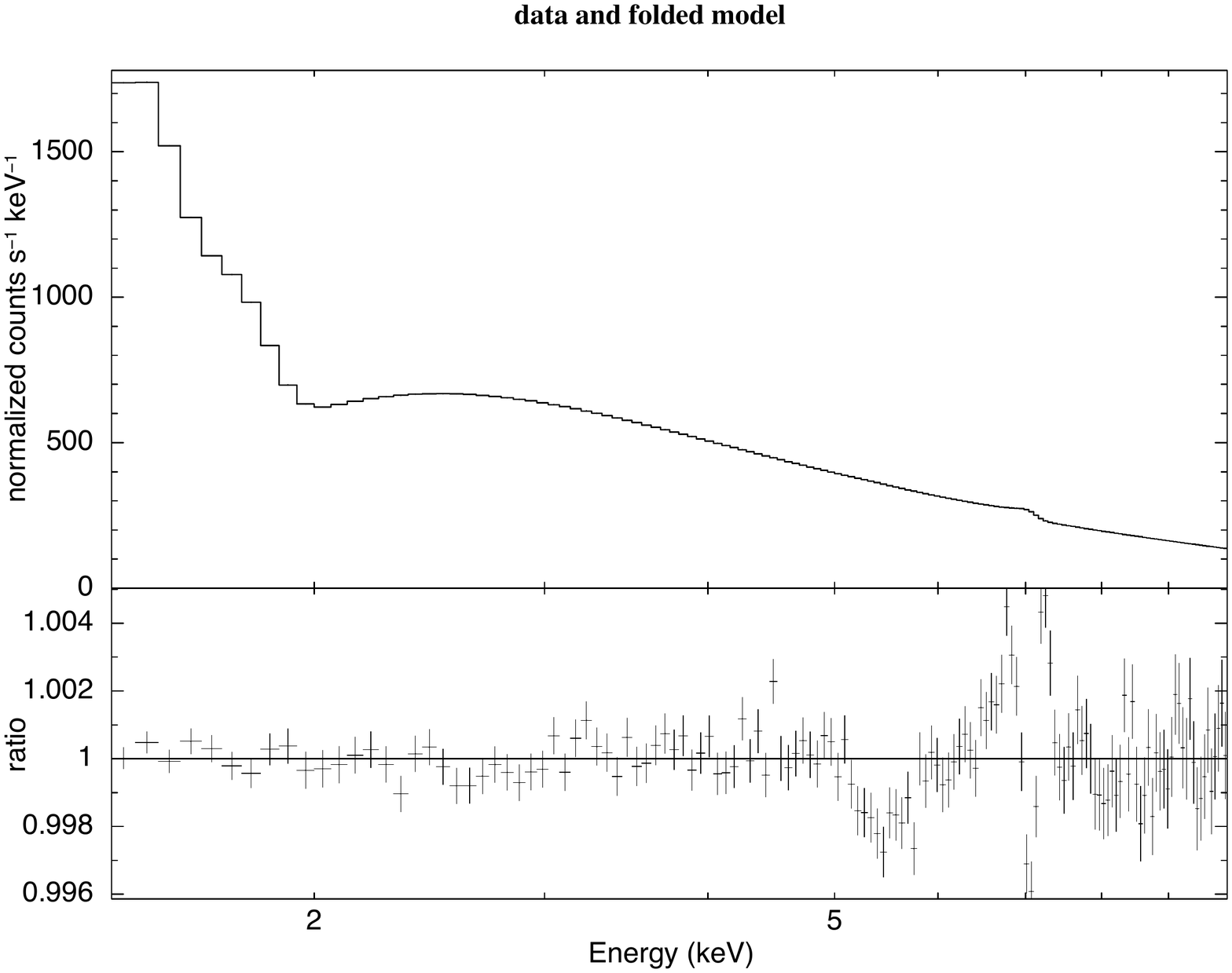} \\
\includegraphics[type=pdf,ext=.pdf,read=.pdf,width=9cm]{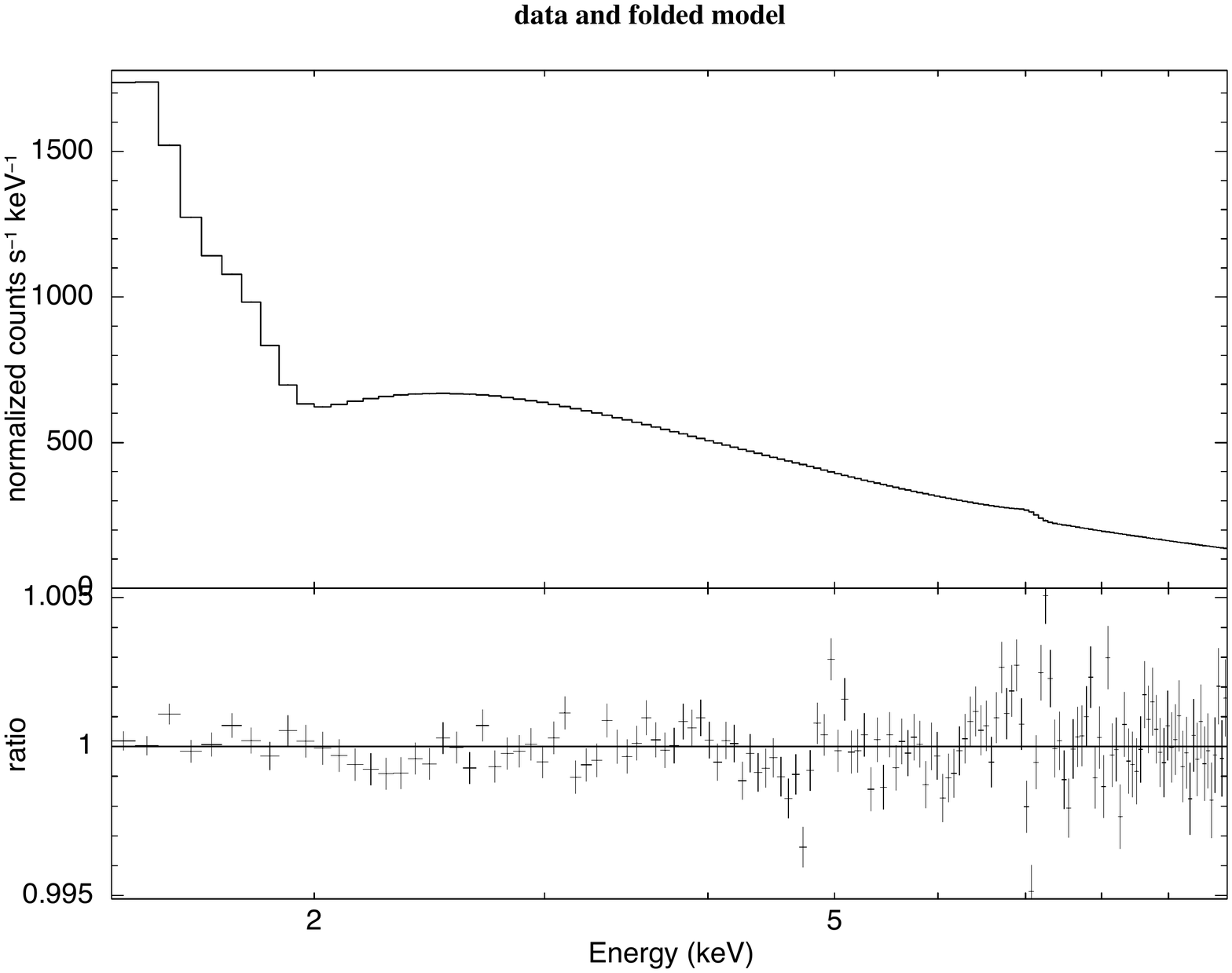}
\hspace{-0.5cm}
\includegraphics[type=pdf,ext=.pdf,read=.pdf,width=9cm]{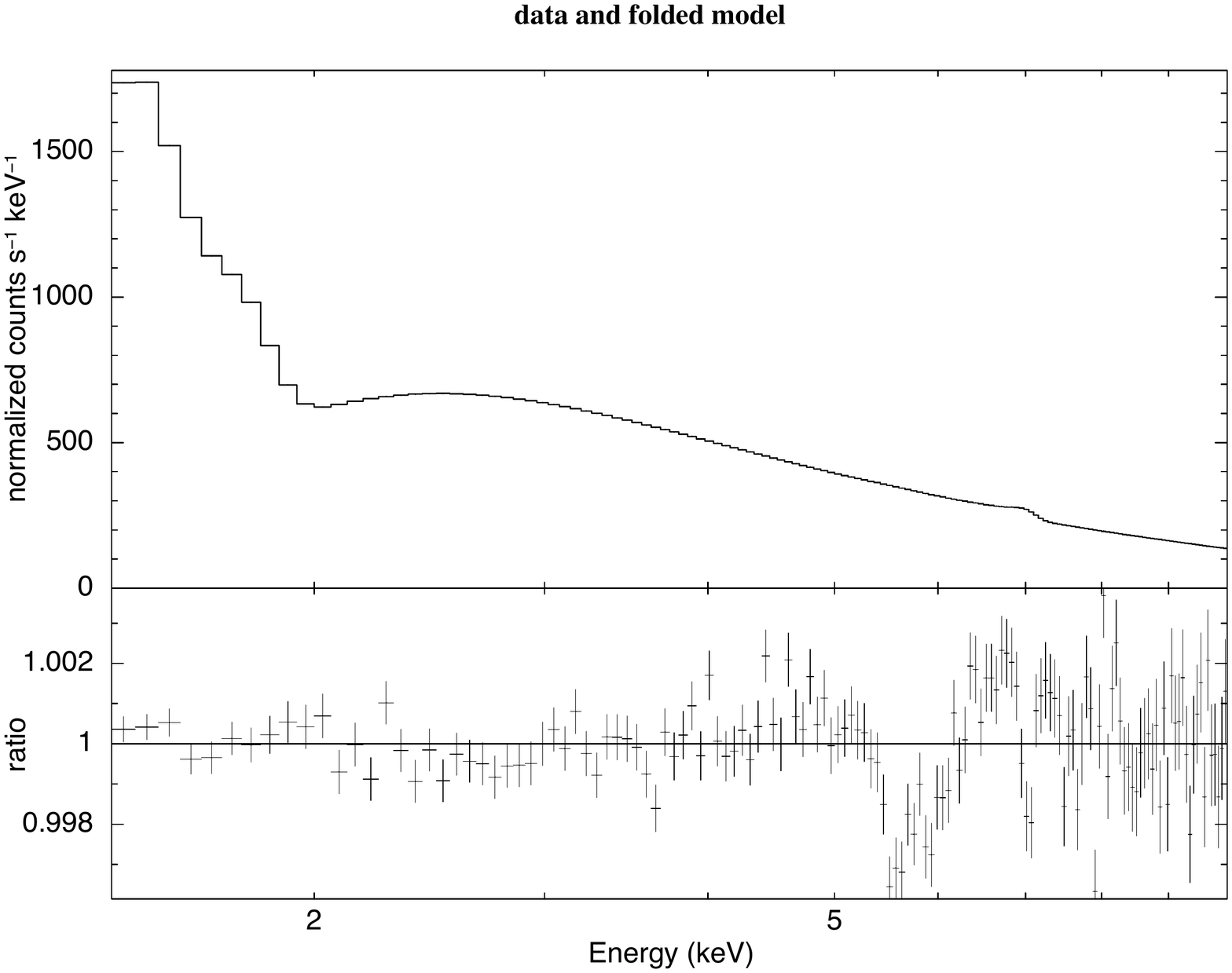} \\
\includegraphics[type=pdf,ext=.pdf,read=.pdf,width=9cm]{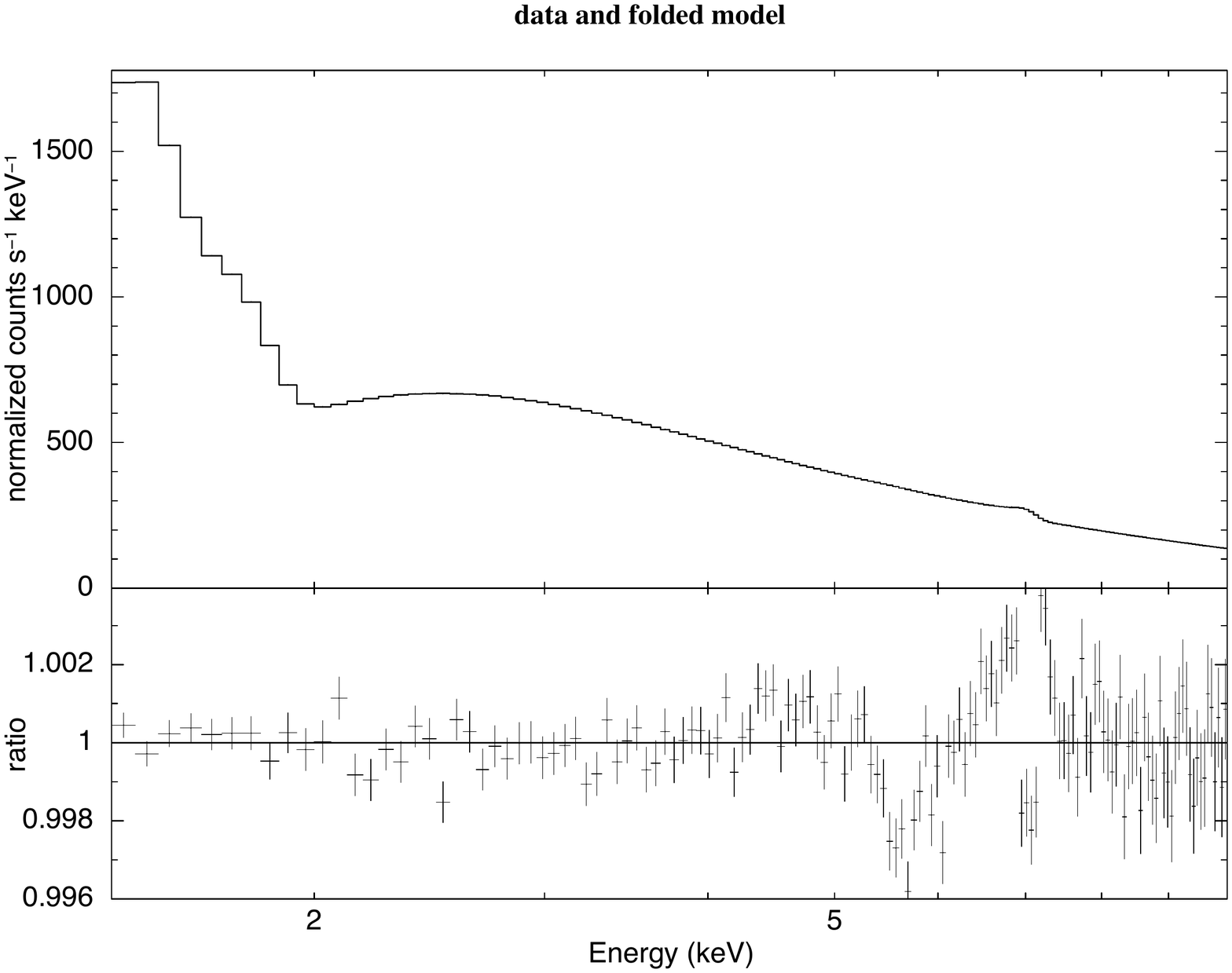}
\hspace{-0.5cm}
\includegraphics[type=pdf,ext=.pdf,read=.pdf,width=9cm]{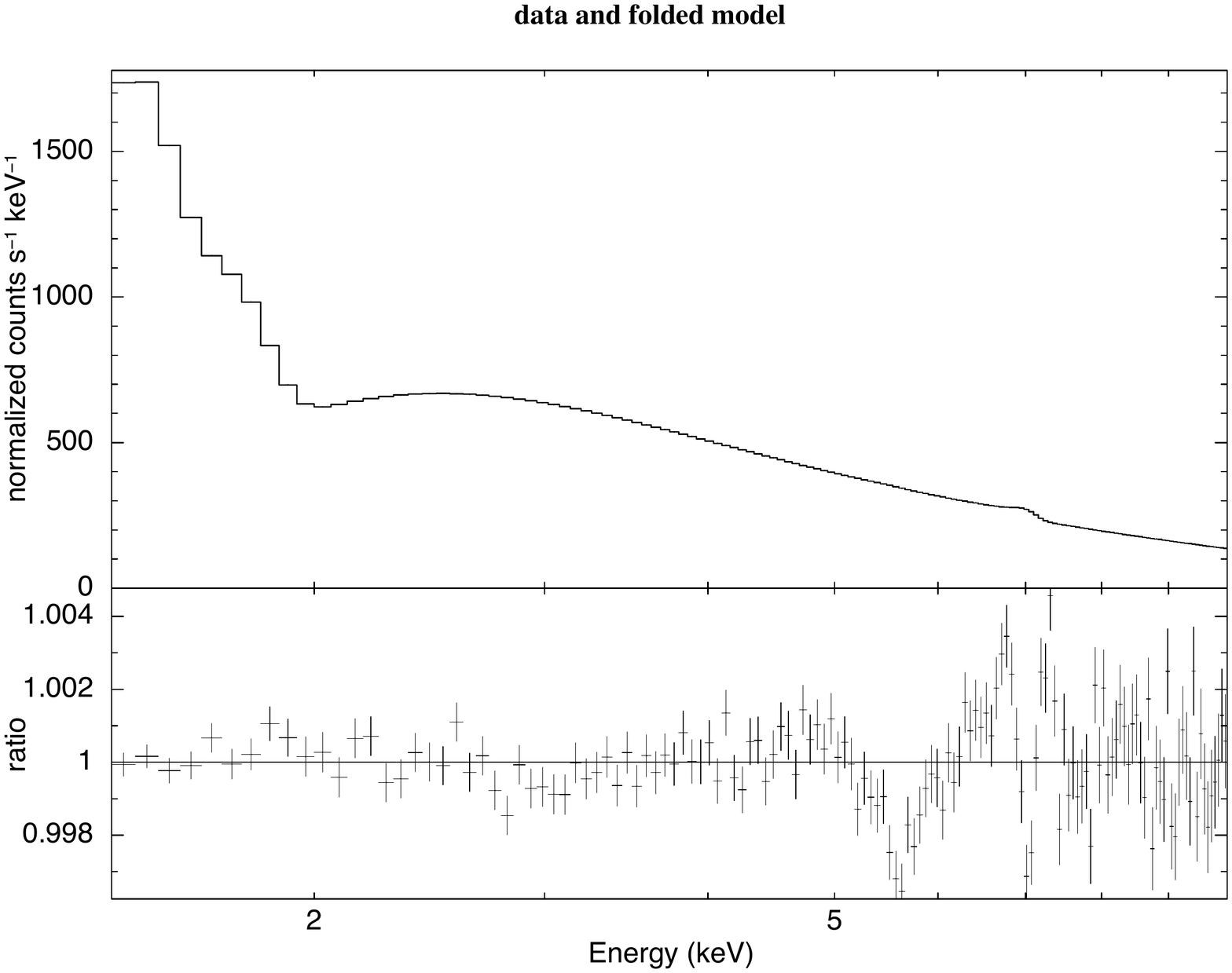}
\end{center}
\vspace{-0.5cm}
\caption{As in Fig.~\ref{f-nustar-b} for the simulations with LAD/eXTP. \label{f-lad-b}}
\end{figure*}

The results of our simulations are shown in Figs.~\ref{f-nustar-a} and \ref{f-nustar-b} for NuSTAR and in Figs.~\ref{f-lad-a} and \ref{f-lad-b} for LAD/eXTP. The values of the reduced $\chi^2$ for the best fit of any observation are reported in the fifth and sixth column of Tab.~\ref{tab}, respectively for NuSTAR and LAD/eXTP.

\section{Discussion \label{s-dis}}

The results of the simulations with NuSTAR are easy to interpret. The 12~spacetimes of our sample are definitively too similar to the Kerr background to see any difference. In all the simulations, the reduced $\chi^2$ of the best-fit is always close to 1 (fifth column in Tab.~\ref{tab}). If we look at the bottom quadrant in the panels in Figs.~\ref{f-nustar-a} and \ref{f-nustar-b}, we do not see any unresolved feature; that is, the Kerr model can fit well the simulated data. Note also that we are considering a particularly bright source, with a relatively long exposure time, and the simulated spectrum is quite simple, with a power-law and an iron line. Moreover the emissivity profile is a simple power-law. All these ingredients should help to test the metric in the strong gravity region. Despite that, we do not see any appreciable difference.

The simulations with LAD/eXTP show that, potentially, we can distinguish the black holes in EdGB gravity from those in Einstein's gravity. However, within our preliminary study with a simplified model it would be dangerous to claim that this is indeed the case. As we have already stressed for the simulations of NuSTAR, the simulated observations should be quite favorable to identify differences with respect to Kerr spacetime. For real data, it may be more difficult. The whole reflection spectrum has additional parameters to be fitted by the data, and the spectrum of the source is more complex. Parameter degeneracy is the main problem in this kind of tests.

It is remarkable that in the simulations with LAD/eXTP we see the same (or very similar) unresolved feature for all the simulations, with the exception of solutions~1 and 9 in which the unresolved feature is present but quite weak. The Kerr model seems to provide a shortage of counts between~5 and 6~keV, and an excess of counts around 7~keV. It seems to be a characteristic of these black holes with respect to the iron line calculated in the Kerr metric, at least when the intensity profile is modeled with a power-law $1/r^q$. If a similar property remains even with more sophisticated models, it may be the signature to look for in order to test EdGB gravity.

\section{Concluding remarks \label{s-con}}

EdGB gravity is one of the simplest string-inspired 4-dimensional models with higher curvature terms. In addition to having a number of appealing theoretical features, its rotating black hole solutions are known numerically, making this theory quite an exceptional case in the panorama of alternative theories of gravity. In the present paper we have studied the possibilities of distinguishing the Kerr black holes of Einstein's gravity from the black holes in EdGB gravity with present and future X-ray missions from the observation of the reflection spectrum, the so called iron line method.

As a preliminary analysis, our study is based on a set of simulations of the spectrum of accreting black holes in EdGB gravity. The simulations are then fitted with a Kerr model. If we obtain a good fit, the observation cannot distinguish a Kerr black hole of Einstein's gravity from a black hole in EdGB gravity. If it is not possible to find a good fit, the two classes of objects can be distinguished and we can constrain the fundamental constants in EdGB gravity.

As an example of a current X-ray mission, we have considered NuSTAR. All the simulations with NuSTAR can be fitted well with a Kerr model. Our conclusion is that current X-ray mission cannot test EdGB gravity from the observations of black holes. Since all the fits are good, it seems we cannot put on any constraint with this technique at the moment. Note also that we are considering a bright source and a simple spectrum, which should help somewhat to distinguish these metrics from the Kerr spacetime, so we expect that our conclusion is quite robust.

In order to study the opportunities offered by the next generation of X-ray missions, we have simulated observations of the same set of numerical metrics with LAD/eXTP. Now the reduced $\chi^2$ of the best-fit with the Kerr model is not so close to 1 and, more importantly, from the ratio between the simulated data and the best-fit we see a more or less prominent unresolved feature (but in some simulations it is not so clear), which is a hint that the fitting model is wrong; that is, the iron line of a black hole in EdGB gravity cannot be fitted with that calculated in the Kerr metric. It is also remarkable that we find the same behavior in all the simulations: a shortage of counts between 5 and 6~keV and an excess of count around 7~keV.

As a final issue we would like to 
connect the results here with those in \cite{res2}, where the shadow
of EdGB black holes was investigated, and only small deviations
from the shadow of Kerr black holes were found.
The results in this work seem to support the message in \cite{res2},
leading to the conclusion that EdGB BHs appear to be rather difficult 
to constrain with tests based on electromagnetic radiation.


\begin{acknowledgments}
The work of H.Z., M.Z., and C.B. was supported by the NSFC (Grant No.~U1531117) and Fudan University (Grant No.~IDH1512060). C.B. also acknowledges the support from the Alexander von Humboldt Foundation.
B.K., and J.K. gratefully acknowledge support by the DFG
Research Training Group 1620 ``Models of Gravity''
and by the grant FP7, Marie Curie Actions, People,
International Research Staff Exchange Scheme (IRSES-606096).
E.R. was supported by FCT-IF programme and the  European  Union's  Horizon  2020  research  and  innovation  programme  under  the  Marie
Sklodowska-Curie grant agreement No 690904 and by the CIDMA project UID/MAT/04106/2013.
\end{acknowledgments}


\end{document}